\documentclass[12pt,oneside,english]{amsart}
\usepackage[T1]{fontenc}
\usepackage[utf8]{inputenc}
\usepackage[letterpaper]{geometry}
\usepackage{float}
\usepackage{booktabs}
\usepackage{amstext}
\usepackage{amsthm}
\usepackage{amssymb}
\usepackage{graphicx}
\usepackage{setspace}
\usepackage{xr}

\usepackage{verbatim}
\usepackage[authoryear]{natbib}
\makeatletter

\providecommand{\tabularnewline}{\\}
\floatstyle{ruled}
\newfloat{algorithm}{tbp}{loa}
\providecommand{\algorithmname}{Algorithm}
\floatname{algorithm}{\protect\algorithmname}

\numberwithin{equation}{section}
\numberwithin{figure}{section}
\theoremstyle{plain}
\newtheorem{thm}{\protect\theoremname}
\include{def}
\def\dispmuskip{\thinmuskip= 3mu plus 0mu minus 2mu \medmuskip=  4mu plus 2mu minus 2mu \thickmuskip=5mu plus 5mu minus 2mu}
\def\textmuskip{\thinmuskip= 0mu                    \medmuskip=  1mu plus 1mu minus 1mu \thickmuskip=2mu plus 3mu minus 1mu}
\def\beq{\dispmuskip\begin{equation}}    \def\eeq{\end{equation}\textmuskip}
\def\beqn{\dispmuskip\begin{displaymath}}\def\eeqn{\end{displaymath}\textmuskip}
\def\bea{\dispmuskip\begin{eqnarray}}    \def\eea{\end{eqnarray}\textmuskip}
\def\bean{\dispmuskip\begin{eqnarray*}}  \def\eean{\end{eqnarray*}\textmuskip}

\usepackage{chngcntr}

\theoremstyle{plain}

\newtheorem{lemma}{Lemma}
\newtheorem{corollary}{Corollary}
\newtheorem{rem}{Remark}

\theoremstyle{plain}
\newtheorem{ass}{Assumption}

\numberwithin{equation}{section}

\usepackage[within=none]{caption}

\usepackage{subfigure}
\usepackage{babel}
\usepackage{amsmath}
\usepackage{mathtools}
\usepackage{bbm}
\usepackage{relsize}

\usepackage{algpseudocode}

\newcommand{\ov}{\overline}

\usepackage [displaymath,mathlines]{lineno}

\def\wt{\widetilde}
\def\wh{\widehat}
\def\ov{\overline}
\def\ul{\underline}
\def\transp{\tiny T}
\def\E{\mathrm{E}}
\def\V{\mathrm{Var}}
\makeatother

\usepackage{babel}
\providecommand{\theoremname}{Theorem}

\begin{document}

\title[Speeding Up MCMC]{Speeding up MCMC by Efficient Data Subsampling}

\author{Matias Quiroz, Robert Kohn, Mattias Villani and Minh-Ngoc Tran}

\thanks{Quiroz and Kohn: \emph{Australian School of Business, University
of New South Wales}. Villani: \textit{Division of Statistics and Machine
Learning, Department of Computer and Information Science, Link\"{o}ping
University}. Tran:\emph{ Discipline of Business Analytics, University
of Sydney.}}
\begin{abstract}
We propose Subsampling MCMC, a Markov Chain Monte Carlo (MCMC) framework
where the likelihood function for $n$ observations is estimated from
a random subset of $m$ observations. We introduce a highly efficient
unbiased estimator of the log-likelihood based on control variates,
such that the computing cost is much smaller than that of the full
log-likelihood in standard MCMC. The likelihood estimate is bias-corrected
and used in two dependent pseudo-marginal algorithms to sample from
a perturbed posterior, for which we derive the asymptotic error with
respect to $n$ and $m$, respectively. We propose a practical
estimator of the error and show that the error is negligible even
for a very small $m$ in our applications. We demonstrate that Subsampling
MCMC is substantially more efficient than standard MCMC in terms of
sampling efficiency for a given computational budget, and that it
outperforms other subsampling methods for MCMC proposed in the literature.

\newpage{}
\end{abstract}

\maketitle

\section{Introduction\label{sec:Introduction}}
Bayesian methods became much more popular after 1990 due to advances in computer technology and the introduction
of powerful simulation algorithms such as Markov Chain Monte Carlo
(MCMC) \citep{gelfand1990sampling}. However, posterior sampling with
MCMC is still time-consuming and there is an increasing awareness that new
scalable algorithms are necessary for MCMC to remain an attractive
choice for inference in data sets with a large number of observations.

\subsection*{Scalable MCMC}Current research on scalable MCMC algorithms belongs to two major
groups. The first group employs parallelism through the typical MapReduce
scheme \citep{dean2008mapreduce} by partitioning the data and computing
separate subposteriors for each partition in a parallel and distributed manner, see for example \citealp{scott2013bayes,neiswanger2013asymptotically,wang2013parallel,minsker2014scalable,nemeth2016merging}.
Our approach belongs to the second group of methods that use a subsample
of the data in each MCMC iteration to speed up the algorithm, which
we refer to as Subsampling MCMC, see \citet{korattikara2014austerity,bardenet2014towards,maclaurin2014firefly,bardenet2015markov,liuexact2015fireflycousin}.
Section~\ref{subsec:Experiment-2} compares these approaches against our methods. See \citet{bardenet2015markov} for an excellent review of these methods and a broad overview of the problem in general.

\subsection*{Pseudo-marginal MCMC} For models where the likelihood cannot be computed analytically (intractable likelihood)
\citet{beaumont2003estimation} proposes estimating
the likelihood unbiasedly and running a Metropolis-Hastings (MH) algorithm on an extended space,
which also includes the auxiliary random variables used to form the likelihood estimate.
\citet{andrieu2009pseudo} call this a Pseudo-Marginal (PM) approach and prove that PM methods target the true posterior density if
the likelihood estimator is unbiased and almost surely positive.

\subsection*{Our contribution} Our article uses the PM framework where at each iteration the log-likelihood from $n$ observations is estimated unbiasedly from a random subset with $m\ll n$ observations,
and the resulting likelihood estimate is then bias corrected to obtain an approximately unbiased estimate of the likelihood. The reason for doing subsampling is because we consider problems
where computing the full likelihood is feasible but inordinately expensive.
This leads to a pseudo marginal sampling scheme targeting a slightly perturbed posterior which mixes well because we use control variates to significantly reduce the variability in the log-likelihood estimate and a correlated pseudo marginal scheme to improve the acceptance probability in the Metropolis-Hastings as discussed below. The control variates are crucial for reducing the variance of the likelihood estimate, and we propose a mixed strategy involving two types of approximations of the log-likelihood contributions of individual data items: i) Taylor expansion around a reference value in parameter space (parameter expanded control variates) \citep{bardenet2015markov} and ii) Taylor expansion around the nearest centroid in data space (data expanded control variates).

We show that by taking $m = O(n^{\frac12})$, the total variation norm of the error in the perturbed posterior is $O(n^{-2})$ if we have access to the Maximum Likelihood Estimate (MLE) based on all data for constructing the control variates, or $O(n^{-\frac12})$ if the MLE is based on a subset with $\tilde n = O(n^\frac12)$ observations. We further show heuristically and also empirically that the proportional error in the perturbed posterior is considerably smaller in regions of high posterior concentration.
We also provide feasible estimators of the proportional error in the perturbed posterior and show empirically
that this error is extremely small in our examples. Finally, our pseudo marginal scheme is
straightforward to implement and tune.

\subsection*{Variance of the likelihood estimator and scalability}
The variance of the log of the estimated likelihood is crucial for
the performance of PM algorithms: a large variance can easily produce
extreme over-estimates of the likelihood and cause the Markov chain
to get stuck for long periods. Conversely, a too precise likelihood
estimator might be unnecessarily costly. \citet{pitt2012some}, \citet{doucet2012efficient}
and \citet{sherlock2013efficiency} analyze the variance of the log of the likelihood
estimator that maximizes the number of effective
draws per unit of computing time. They conclude that the optimal number
of particles $m$ should be such that this variance is around
$1$. Moreover, $m=O(n)$ is required to obtain the optimal value
of the variance.

Obtaining unbiased likelihood estimators
with low variability from subsampling is a major challenge, and previous
attempts have failed to produce an MCMC sampler that does not get
stuck \citep{korattikara2014austerity,bardenet2015markov}. Moreover,
ensuring that the unbiased likelihood estimator is also positive was
shown by \citet{jacob2015nonnegative} to be possible under assumptions
that can only be satisfied by sampling the full data set \citep{bardenet2015markov}.

It is now recognised that it is the variance of the difference in the logs of the likelihood estimators
at the current and proposed values of the parameters that must be controlled. In the standard PM this is equivalent
to controlling the variance of the log of the estimated likelihood.

Recent advances in PM algorithms correlate or block the random numbers used to form the estimates of the likelihood in the MH ratio at the current and proposed values of the parameters \citep[see][respectively]{deligiannidis2015correlated,tran2016block}.
\cite{deligiannidis2015correlated} show that this makes it possible to target a variance of the log estimated likelihood that is much larger than one, and the optimal variance
   can be obtained with $m = O(n^{1/2}) $. \cite{dahlin2015correlated}
also introduces the correlated PM but their paper does not contain any analytic or optimality results. \cite{tran2016block} give an alternative derivation of this result and generalize it to the case where the likelihood is estimated by randomized quasi Monte Carlo. Our article introduces both the correlated and block correlated PM approaches to data subsampling.

\subsection*{Related approaches using our subsampling methods} The subsampling methods and theory proposed here have already found applications in several recently proposed algorithms.

\citet{quiroz2016exact} use the insights and methods of our article (control variates and correlated and block PM for subsampling) to obtain unbiased estimates of posterior expectations of functions of the parameters. The method uses a version of the unbiased, but possibly negative, Poisson estimator \citep{wagner1988monte} of the likelihood and runs a PM algorithm based on the absolute value of this estimator. The resulting iterates are subsequently used in an importance sampling scheme following \citet{lyne2015russian} to obtain simulation consistent posterior expectations of functions of the parameters. Although exact, this approach has some drawbacks compared to the approach proposed here. First, it does not automatically produce an estimate of the posterior distribution of a function of the parameters because it is not an MCMC approach, and hence it is infeasible in practice to obtain credible regions with it. Second, the approach in \citet{quiroz2016exact} is more sensitive to the variance of the likelihood estimator than the approach presented here, in the following way. Let $\wh L_{exact}$ be the unbiased but possibly negative likelihood estimator in \citet{quiroz2016exact} and let $\wh L_{approx}$ be the perturbed likelihood estimate considered in our article. We can then show that $\V (\log |\wh L_{exact}|) \approx \exp(\V (\log \wh L_{approx})) - 1$ for the same computational cost. This means that the two variances are approximately equal if $\V (\wh L_{approx}) \ll 1$, but that $\V (\log |\wh L_{exact}|)$ can be much larger than $\V (\log \wh L_{approx})$ if $\V (\wh L_{approx}) \gg 1$.

\citet{quiroz2015delayed} apply the framework, methodology and theory of a previous version of our paper to propose a delayed acceptance subsampling scheme which they implement using the data expanded control variates. Unlike Theorem~\ref{thm:maintheorem} and Corollary~\ref{corr: cv implications} of our article, there are no theoretical or empirical results of how the parameter expanded control variates affect the error in the perturbed posterior.

\subsection*{Article outline}
The paper is organized as follows. Section~\ref{sec:Sampling-based-Log-l-Estimators}
introduces the general likelihood estimator and derives some important properties.
Section~\ref{sec:theory} outlines the subsampling MCMC algorithm
and its theoretical framework, including results on the accuracy of
the perturbed posterior.
Section~\ref{sec:Application}
studies empirically our proposed methodology and shows that it outperforms
both standard (non-subsampling) MCMC and other subsampling approaches. There is online supplementary material to the
paper. We refer to pages, sections, etc in the supplement as Page S1, Section~S1, etc.
Section~\ref{AppendixTop:proxies} contains implementation details, Section~\ref{AppendixTop:Proofs} contains some
proofs and Section~\ref{app:glms} shows how our theory applies to generalized linear models.

\section{Sampling-based Log-likelihood Estimators\label{sec:Sampling-based-Log-l-Estimators}}

\subsection{A log-likelihood estimator based on simple random sampling with efficient
control variates\label{subsec:DE}}

Let $\left\{ y_{i},x_{i}\right\} _{i=1}^{n}$ denote the data, where
$y$ is a response vector and $x$ is a vector of covariates. Let
$\theta\in\Theta$ be a $p$-dimensional vector of parameters. Given
conditionally independent observations we have the usual decomposition
of the log-likelihood
\begin{equation}
\ell_{(n)}(\theta)\coloneqq\sum_{i=\text{1}}^{n}\ell_{i}(\theta),\quad\text{where }\ell_{i}(\theta)\coloneqq\log p(y_{i}|\theta,x_{i})\label{eq:cross_sec_likelihoods}
\end{equation}
is the \emph{log-likelihood contribution} of the $i$th observation.
For any given $\theta$, \eqref{eq:cross_sec_likelihoods} is a sum
of a finite number of elements and estimating it is equivalent to
the classical survey sampling problem of estimating a population total.
See \citet{sarndal2003model} for an introduction. We assume in \eqref{eq:cross_sec_likelihoods}
that the log-likelihood decomposes as a sum of terms where each term
depends on a unique piece of data information. This applies to longitudinal
problems where $\ell_{i}(\theta)$ is the log joint density of all
measurements on the $i$th subject, and we sample subjects rather
than individual observations. It also applies to certain time-series
problems such as $\mathrm{AR}(l)$ processes, where the sample elements
become $(y_{t},\dots,y_{t-l})$, for $t=l+1,\dots,n$. Our examples
in Section~\ref{sec:Application} use independent identically distributed
(iid) observations and time series data.

Estimating \eqref{eq:cross_sec_likelihoods} using Simple Random
Sampling (SRS), where any $\ell_{i}(\theta)$ is included with the
same probability generally results in a very large variance.
Intuitively, since some $\ell_{i}(\theta)$ contribute significantly
more to the sum in \eqref{eq:cross_sec_likelihoods} they should be
included in the sample with a larger probability, using so called
Probability Proportional-to-Size (PPS) sampling. However, this requires
each of the $n$ sampling probabilities to be proportional to a measure
of their size. Evaluating $n$ size measures is likely to defeat the
purpose of subsampling, except when there is a computationally
cheaper proxy than $\ell_{i}(\theta)$ that can be utilized instead.
Alternatively, one can make the $\{\ell_{i}(\theta)\}_{i=1}^{n}$
more homogeneous by using control variates so that the population
elements are roughly of the same size and SRS is then expected to
be efficient. Our article focuses on this case and proposes efficient
control variates $q_{i,n}(\theta)$ such that the computational cost
of the estimator is substantially less than $O(n)$. The dependence
on $n$ is due to $q_{i,n}(\theta)$ being an approximation of $\ell_{i}(\theta)$,
which typically improves as more data is available as we will discuss
in detail later.

Define the differences $d_{i,n}(\theta)\coloneqq\ell_{i}(\theta)-q_{i,n}(\theta)$
and let
\[
\mu_{d,n}(\theta)\coloneqq\frac{1}{n}\sum_{i=1}^{n}d_{i,n}(\theta)\quad\text{and }\sigma_{d,n}^{2}(\theta)\coloneqq\frac{\sum_{i=1}^{n}\left(d_{i,n}(\theta)-\mu_{d,n}(\theta)\right)^{2}}{n}
\]
be the mean and variance of the finite population $\{d_{i,n}(\theta)\}_{i=1}^{n}$.
Let $u_{1},\dots,u_{m}$ be iid random variables such that $\Pr(u=k)=1/n$
for $k=1,\dots,n$. The Difference Estimator (DE, \citealp{sarndal2003model})
of $\ell_{(n)}(\theta)$ in \eqref{eq:cross_sec_likelihoods} is
\begin{equation}
\widehat{\ell}_{(m,n)}(\theta)\coloneqq q_{(n)}(\theta)+n\widehat{\mu}_{d,n}(\theta),\quad\widehat{\mu}_{d,n}(\theta)\coloneqq\frac{1}{m}\sum_{i=1}^{m}d_{u_{i},n}(\theta), \label{eq:DE_Estimator}
\end{equation}
with $q_{(n)}(\theta)\coloneqq\sum_{i=1}^{n}q_{i,n}(\theta)$. It
is straightforward to use unequal sampling probabilities with the
DE, but the sampling probabilities need to be evaluated for every
observation, which can be costly. The following lemma gives some basic
properties of the DE estimator. Its proof is in Appendix~\ref{AppendixTop:Proofs}.
\renewcommand{\labelenumi}{(\roman{enumi}).}
\begin{lemma}\label{lem:Unbiased_and_variances}Suppose
that $\widehat{\ell}_{(m,n)}(\theta)$ is the estimator of $\ell_{(n)}(\theta)=\ell(\theta)$
given by \eqref{eq:DE_Estimator}. Then, for each $\theta$,
\begin{enumerate}
\item $\E[\widehat{\mu}_{d,n}(\theta)]=\mu_{d,n}(\theta)$.
\item
\[
\E\left[\widehat{\ell}_{(m,n)}(\theta)\right]=\ell_{(n)}(\theta)\quad\text{and}\quad\sigma^{2}_{LL,m,n}(\theta):=\V\left[\widehat{\ell}_{(m,n)}(\theta)\right]=\frac{n^{2}\sigma_{d,n}^{2}(\theta)}{m}.
\]
\item $\widehat{\ell}_{(m,n)}(\theta)$ is asymptotically normal when $m\rightarrow\infty$
for fixed $n$ and $\sigma_{d,n}^{2}(\theta)<\infty$, or when both
$m,n\rightarrow\infty$ with $m=O(n^{\alpha})$ for $\alpha>0$ and $\sigma^{3}_{d,n}(\theta)<\infty$.
\end{enumerate}
\end{lemma}The assumptions of finite $\sigma_{d,n}^{2}(\theta)$
and $\sigma_{d,n}^{3}(\theta)$ in Lemma \ref{lem:Unbiased_and_variances}
part (iii) are non-restrictive because the random variables
are discrete with a finite sample space: they are satisfied for any
control variates that are finite. We use the following estimate of
$\sigma_{d,n}^{2}(\theta)$
\[
\widehat{\sigma}{}_{d,n}^{2}(\theta)\coloneqq\frac{\sum_{i=1}^{m}\left(d_{u_{i},n}(\theta)-\widehat{\mu}_{d,n}(\theta)\right)^{2}}{m}.
\]
We also define the higher order central moments
$$\varphi_{d,n}^{(b)}(\theta) \coloneqq \E[(d_{u_{i},n}(\theta)-\mu_{d,n}(\theta))^{b}]=\sum_{i=1}^{n}(d_{i,n}(\theta)-\mu_{d,n}(\theta))^{b}/n \quad \text{for } b \geq 1,$$
and the corresponding standardized quantities
$\Psi_{d,n}^{(b)}(\theta)\coloneqq\varphi_{d,n}^{(b)}(\theta)/\sigma_{d,n}^{b}(\theta).$

\subsection{Control variates for variance reduction and optimal subsample size\label{subsec:Control-variates}}

We will now show that the variance reduction from control variates has a dramatic effect on how the subsample size $m$
relates to the sample size $n$. The theory on how to choose the number of particles in PM in \cite{pitt2012some} and
\cite{doucet2012efficient} is based on minimization of the computational cost of obtaining a single posterior draw that corresponds to an iid draw, see e.g. \citealp{pitt2012some,doucet2012efficient}. This theory assumes that the likelihood is estimated directly, rather than indirectly via a bias-corrected log-likelihood estimator as proposed here. The relevant cost for evaluating the likelihood estimator in \cite{pitt2012some} and \cite{doucet2012efficient} can therefore be argued to be inversely proportional to variance of the log of the likelihood estimator, and the optimal number of particles or subsampled units $m$ targets a variance of the log of the likelihood estimator around one. In our approach the estimation effort is instead spent on estimating the log-likelihood. The relevant computational cost is therefore inversely proportional to $\sigma^{2}_{LL,m,n}$ and the optimal $m$ targets a $\sigma^{2}_{LL,m,n}$ of $O(1)$. See
Section~\ref{subsec:Choosing-the-sampling} for more details.

Lemma \ref{lem:m_and_n_asymptotics} below details the asymptotic behavior of $\sigma^{2}_{LL,m,n}$ using the definition
\begin{equation}
a_{n}(\theta)\coloneqq2\max_{i=1,\dots,n}\left|d_{i,n}(\theta)\right|.\label{eq:a_n_sequence}
\end{equation}

The proof of the following lemma is straightforward and therefore omitted. All terms in the lemma depend on $\theta$.
\begin{lemma}\label{lem:m_and_n_asymptotics}For each $\theta\in\Theta$,
\begin{enumerate}
\item[(i)] $\sigma_{d,n}^{b}=O(a_{n}^{b})$ for $b \geq 1$. In particular, $\sigma_{d,n}^{2}=O(a_{n}^{2})$.
\item[(ii)] $\sigma_{LL,m,n}^{2}=\frac{n^{2}O(a_{n}^{2})}{m}$.
\item[(iii)] $\varphi_{d,n}^{(b)}=O(a^b_n)$ and $\Psi_{d,n}^{(b)} = O(1)$.
\end{enumerate}
\end{lemma}
Part (ii) of Lemma \ref{lem:m_and_n_asymptotics} shows that keeping the variance of the log-likelihood estimate bounded
as a function of $n$ requires that $\frac{n^{2}O(a_{n}^{2})}{m}=O(1)$.
This highlights the importance of the variance reduction: SRS without
control variates scales poorly because $O(a_{n}^{2})=O(1)$ and so $m=O(n^{2})$ is optimal.
Conversely, with control variates that improve as, say $d_{i,n}=O(n^{-\alpha})$
with $\alpha\geq0$, we have $O(a_{n}^{2})=O(n^{-2\alpha})$ and $m=O(n^{2(1-\alpha)})$ is optimal. Lemma \ref{lem:m_and_n_asymptotics} also shows the asymptotic properties of the central moments, which are useful for our derivation of the perturbed target in Section~\ref{subsec:A-heuristic-to}.

\subsection{Computational complexity\label{subsec:Computational-complexity}}

The difference estimator in \eqref{eq:DE_Estimator}
requires computing $q_{(n)}(\theta) = \sum_{i=1}^n q_{i,n}(\theta) $ in every MCMC iteration, i.e.,
it requires computing the control variates $q_{i,n}(\theta) $ for $i=1, \dots, n $.
We now explore specific choices of $q_{i,n}$ that
allow us to compute $\sum_{i=1}^{n}q_{i,n}(\theta)$ using substantially
less evaluations than $n$. Denote the Computational Cost (CC) for
the standard MH without subsampling which evaluates $\ell_{(n)}\coloneqq\sum_{i=1}^{n}\ell_{i}$
by $
\mathrm{CC}[\ell_{(n)}(\theta)]\coloneqq n\cdot c_{\ell},
$
where $c_{\ell}$ is the cost of evaluating a single log-likelihood
contribution (assuming the cost is the same for all $i$). For the
difference estimator in \eqref{eq:DE_Estimator}, we have
\[
\mathrm{CC}\left[\widehat{\ell}_{(m,n)}(\theta)\right]\coloneqq n\cdot c_{q}+m\cdot c_{\ell},
\]
where $c_{q}$ is the cost of computing a control variate. We now
briefly describe two particular control variates that reduce the first
term $n\cdot c_{q}$. Appendix \ref{AppendixTop:proxies} gives
 implementation details.

First, consider the control variates in \citet{bardenet2015markov}
who propose using a second order Taylor expansion of each $\ell_{i}(\theta)$
around some reference value $\theta^{\star}_n$, e.g. the maximum likelihood
estimate. This reduces the complexity from $n$ evaluations to a single
one (similar to sufficient statistics for a normal model because $q_{i,n}(\theta)$
is quadratic in $\theta$). As noted by \citet{bardenet2015markov},
this control variate can be a poor approximation of $\ell_{i}(\theta)$
whenever the algorithm proposes a $\theta$ that is not near to $\theta^{\star}_n$,
or when there is no access to a reasonable $\theta^{\star}_n$.

Second, we propose a new control variate
which is based on clustering the data $\{z_{i}=(y_{i},x_{i})\}_{i=1}^{n}$
into $K$ clusters that are kept fixed, and is independent of $\theta^{\star}_n$.
 At a given MCMC iteration,
we compute the exact log-likelihood contributions at all $K$ centroids
and use a second order Taylor expansion with respect to $z_{i}$ at
the centroid $z^{c}$ as a local approximation of $\ell_{i}$ around
each centroid. This allows us to compute $\sum_{i=1}^{n}q_{i,n}(\theta)$
by evaluating quantities computed at the $K$ centroids (similar to
sufficient statistics for a normal model because $q_{i,n}(\theta)$
is now quadratic in $z$). The cost of the resulting estimator is
\begin{equation}
\mathrm{CC}\left[\widehat{\ell}_{(m,n)}(\theta)\right]=K\cdot c_{q}+m\cdot c_{\ell},\label{eq:ComputingCostDE_ourProxies}
\end{equation}
where typically $K\ll n$.

We refer to the control variate that uses a Taylor expansion
with respect to $\theta$ as \emph{parameter expanded}, and the control variate
type that Taylor expands with respect to $z$ as \emph{data expanded}.

\subsection{Asymptotic properties of the control variates\label{subsec:Asymptotic-properties}}
\subsubsection{Data expanded control variates}
To derive the asymptotic behavior of $a_n(\theta)$ in \eqref{eq:a_n_sequence} for data expanded control variates we bound the remainder term \cite[Appendix A.9]{Hubbard:1999}  \[
\left|d_{i,n}(\theta)\right|\leq O\left(\left(||z-z^{c}||_{1}\right)^{3}\right)= O\left(\epsilon^{3}\right),
\]
where $||\cdot||_{1}$ denotes the $l_{1}$-norm and $\epsilon$ is an input to Algorithm \ref{Alg:Cluster} in Appendix
\ref{AppendixTop:proxies}, which is proportional to the maximum $l_{1}$-distance
between an observation $z$ and its centroid $z^{c}$. If
the numbers of clusters increases with $n$ such that $\epsilon=O(n^{-\zeta})$
for some $\zeta>0$, then $\alpha=3\zeta$ in $d_{i,n}(\theta)=O(n^{-\alpha})$ and hence $a_n(\theta) = O(n^{-3\zeta})$ for this control
variate. Our simulations show that the numbers of clusters needs to
increase rapidly with $n$ to satisfy the error decay ($\zeta > 0$)
when the effective dimension of the data $\tilde{p}$ is large and data are independent across dimensions (not
shown here); these empirical results are supported by Theorem 5.3b in \citet{graf2002rates}
which states that the mean distance in $k$-means clustering between
an observation to its nearest centroid decreases as $O(n^{-1/(\tilde{p}+2)})$
if the number of centroids grows as $o(n^{\tilde{p}/(\tilde{p}+2)})$ for any distribution
with compact support. However, the performance on real data depends on
the extent to which the observed data lies close to a lower-dimensional
manifold, and we have observed good performance in our examples
in Section~\ref{sec:Application}, where $\tilde{p}\leq21$. Nevertheless,
data expanded control variates will eventually suffer from the curse
of dimensionality, and we now turn to the asymptotic properties of parameter expanded control variates.

\subsubsection{Parameter expanded control variates}
\begin{ass} \label{ass: 3rd deriv assum}
Suppose that for each $i$, $\ell_i(\theta)$ is three times differentiable with
\begin{align*}
\max_{j,k,l \in \{1, \dots, p\}} \sup_{\theta \in \Theta} \Biggr \rvert \frac{\partial^3 \ell_i(\theta)}
{\partial \theta_j\partial \theta_k\partial \theta_l}\Biggr \rvert
\end{align*}
bounded.
\end{ass}
We now have the following result, where $|| \cdot ||$ is the $l_2$ norm for the rest of the paper unless stated otherwise. The proof of the lemma is immediate.
\begin{lemma}
Suppose that Assumption~\ref{ass: 3rd deriv assum} holds. Then, $a_n(\theta) = || \theta - \theta_n^\star||^3  O(1)$
\end{lemma}

While the asymptotics for the data expanded covariates are interpreted in a nonstochastic sense ($z$ is nonstochastic) our interpretation here also treats data as nonstochastic, but the parameter as stochastic so that we can utilize the Bernstein-von Mises theorem (BvM). The BvM theorem states that the posterior distribution converges to the normal distribution (in some sense) when the sample size $n \to \infty$. There are probabilistic (stochastic data) and nonstochastic (nonstochastic data) versions of the BvM and we use a version of the latter one due to \cite{Chen:1985}. Treating the data as fixed leads to a better interpretation in our context and is also consistent with a Bayesian interpretation.

\section{Subsampling MCMC Methodology\label{sec:theory}}
\subsection{MCMC with likelihood estimators from data subsampling}

We propose an efficient unbiased estimator $\widehat{\ell}_{(m,n)}(\theta)$
of the log-likelihood and then approximately bias-correct it following
\citet{ceperley1999penalty} \citep[see also][]{nicholls2012coupled}
to obtain the approximately bias-corrected likelihood estimator
\begin{equation}
\widehat{L}{}_{(m,n)}(\theta,u)\coloneqq\exp\left(\widehat{\ell}_{(m,n)}(\theta)-\frac{n^{2}}{2m}\widehat{\sigma}_{d,n}^{2}(\theta)\right),\label{eq:LikelihoodEstimator}
\end{equation}
where $\widehat{\ell}_{(m,n)}(\theta)$ and $\widehat{\sigma}_{d,n}^{2}(\theta)$
are the estimators presented in Section~\ref{subsec:DE}. The form
of \eqref{eq:LikelihoodEstimator} is motivated by the case when $\widehat{\ell}_{(m,n)}\text{\ensuremath{\sim}}\mathcal{N}(\ell_{(n)}(\theta),\sigma_{LL,m,n}^{2}(\theta))$
and $\sigma_{LL,m,n}^{2}$ is known, in which case all bias is removed.
Normality holds asymptotically in both $m$ and $n$ by part (iii)
of Lemma \ref{lem:Unbiased_and_variances}. However, the assumption
of a known variance is unrealistic because the computation requires
the entire data set. The estimator in \eqref{eq:LikelihoodEstimator}
is therefore expected to only be nearly unbiased.

There are four main differences between our approach and \citet{ceperley1999penalty}
and \citet{nicholls2012coupled}. First, our approach is pseudo marginal and takes into account that the log likelihood is estimated using a random subsample at each iteration and is therefore guaranteed to converge to the posterior distribution. Second, we use control variates to decrease the variance of the estimator of the loglikelihood and analyze the effect that these control variates have on the variance of the log of the estimate of the likelihood. Third, we use correlated pseudo marginal schemes to also allow the log of the estimated likelihood to have a large variance. Finally, our convergence rate of the error (Theorem~\ref{thm:maintheorem} below) is $O(n^{-1}m^{-2})$ as opposed to $O(m^{-1})$ in \citet{nicholls2012coupled}.

We now outline how to carry out a pseudo-marginal MH scheme with the approximately
unbiased estimator in \eqref{eq:LikelihoodEstimator} and derive the
asymptotic error in the stationary distribution. Denote the likelihood
by $L_{(n)}(\theta)\coloneqq p(y|\theta)$, let $p_{\Theta}(\theta)$
be the prior and define the marginal likelihood $\overline{L}{}_{(n)}\coloneqq\int L_{(n)}(\theta)p_{\Theta}(\theta)d\theta$.
Then, the posterior is $\pi_{(n)}(\theta)=L_{(n)}(\theta)p_{\Theta}(\theta)/\overline{L}{}_{(n)}$.
Let $p_{U}(u)$ be the distribution of the vector $u$ of auxiliary
variables corresponding to the subset of observations to include when
estimating $L_{(n)}(\theta).$ Let $\widehat{L}{}_{(m,n)}(\theta,u)$,
for fixed $m$ and $n$, be a possibly biased estimator of $L_{(n)}(\theta)$
with expectation
\[
L_{(m,n)}(\theta)=\int\widehat{L}{}_{(m,n)}(\theta,u)p_{U}(u)du.
\]
Define
\begin{equation}
\overline{\pi}_{(m,n)}(\theta,u)\coloneqq\widehat{L}{}_{(m,n)}(\theta,u)p_{U}(u)p_{\Theta}(\theta)/\overline{L}_{(m,n)},\text{ with }\overline{L}_{(m,n)}\coloneqq\int L{}_{(m,n)}(\theta)p_{\Theta}(\theta)d\theta,\label{eq:AugmentedPosterior}
\end{equation}
on the augmented space $(\theta,u)$. It is straightforward to show
that $\overline{\pi}_{(m,n)}(\theta,u)$ is a proper density with
marginal
\begin{eqnarray*}
\overline{\pi}_{(m,n)}(\theta) & = & \int\overline{\pi}_{(m,n)}(\theta,u)du=L_{(m,n)}(\theta)p_{\Theta}(\theta)/\overline{L}_{(m,n)}.
\end{eqnarray*}

The standard PM that targets \eqref{eq:AugmentedPosterior} uses a joint
proposal for $\theta$ and $u$ given by
\[
q_{\Theta,U}(\theta,u|\theta_{c},u_{c})=p_{U}(u)q_{\Theta}(\theta|\theta_{c}),
\]
where $\theta_c$ denotes the current state of the Markov chain. The PM acceptance
probability becomes
\begin{equation}
\alpha=\min\left(1,\frac{\widehat{L}{}_{(m,n)}(\theta_{p},u_{p})p_{\Theta}(\theta_{p})/q_{\Theta}(\theta_{p}|\theta_{c})}{\widehat{L}{}_{(m,n)}(\theta_{c},u_{c})p_{\Theta}(\theta_{c})/q_{\Theta}(\theta_{c}|\theta_{p})}\right).\label{eq:accprobPseudoMCMC}
\end{equation}
This expression is similar to the standard MH acceptance probability, but with
the true likelihood replaced by its estimate. By \citet{andrieu2009pseudo},
the draws of $\theta$ obtained by this MH algorithm have $\overline{\pi}_{(m,n)}(\theta)$
as invariant distribution. If $\widehat{L}{}_{(m,n)}(\theta,u)$ is
an unbiased estimator of $L_{(n)}(\theta)$, then the marginal of
the augmented MCMC scheme above has $\overline{\pi}_{(m,n)}(\theta)=\pi_{(n)}(\theta)$
(the true posterior) as invariant distribution. However, if $\widehat{L}{}_{(m,n)}(\theta,u)$
is biased, the sampler is still valid but has a perturbed marginal
$\overline{\pi}_{(m,n)}(\theta)$.

\subsection{Perturbation analysis - asymptotics}
The discussion in Section~\ref{subsec:Asymptotic-properties} argued that parameter expanded covariates have better asymptotic properties. We therefore state and prove our main theorem on the fractional error in the  perturbed quantities under this choice of control variate.
Let $\pi_{(n)}(\theta) \propto \exp(\ell_{(n)}(\theta)) p_{\Theta}(\theta)$ be the density function of the posterior distribution of $\theta$, where $p_{\Theta}$ is the prior density for $\theta$. Let $\theta^\star_n$ be a mode of $\pi_{(n)}$, and
 \[\Delta_n(\theta):=\frac{\partial^2\log\pi_n(\theta)}{\partial\theta\partial\theta^T}.\]
Denote by $H(a,\delta)=\{\theta\in\Theta:\|\theta-a\|\leq\delta\}$ a neighbourhood of $a$.
We follow \cite{Chen:1985} and make the following assumptions.

\begin{ass}\label{Ass2}
Assume that the following hold
\begin{itemize}
\item[(A1)] ${\partial\log\pi_n(\theta)}/{\partial\theta}|_{\theta=\theta^\star_n}=0$.
\item[(A2)] $\Delta_n(\theta^\star_n)$ is negative definite.
\item[(A3)] $\|\Sigma_n\|_2=O(n^{-1})$, where $\Sigma_n=\big(-\Delta_n(\theta^\star_n)\big)^{-1}$.
\item[(A4)] For any $\epsilon>0$, there exist a $\delta_\epsilon>0$ and an integer $N_{1,\epsilon}$ such that for any $n>N_{1,\epsilon}$ and $\theta\in H(\theta^\star_n,\delta_\epsilon)$, $\Delta_n(\theta)$ exists and satisfies
\[-A(\epsilon)\leq \Delta_n(\theta)\big(\Delta_n(\theta^\star_n)\big)^{-1}-I\leq A(\epsilon)\]
where $A(\epsilon)$ is a positive semidefinite matrix whose largest eigenvalue goes to 0 as $\epsilon\to0$.
\item[(A5)]  For any $\delta>0$, there exists a positive integer $N_{2,\delta}$ and two positive numbers $c$ and $\kappa$ such that
for $n>N_{2,\delta}$ and $\theta\not\in H(\theta^\star_n,\delta)$
\[\frac{\pi_{(n)}(\theta)}{\pi_{(n)}(\theta^\star_n)}<\exp\left(-c\big[(\theta-\theta^\star_n)^T\Sigma_n^{-1}(\theta-\theta^\star_n)\big]^\kappa\right).\]
\end{itemize}
\end{ass}

\cite{Chen:1985} shows that the conditions in Assumption \ref{Ass2} hold in regular exponential families with conjugate priors. His proof carries directly over to Generalized Linear Models in the canonical parametrization, which includes the logistic regression used in the applications in Section~\ref{sec:Application}. This result also generalizes in a straightforward way to the non-canonical case if the link function has continuous third derivative, see
Section~\ref{app:glms} for details.

\begin{thm}\label{thm:maintheorem}
Suppose that we use parameter expanded control variates and assume that the regularity conditions in Assumption \ref{Ass2} are satisfied. Then
\begin{enumerate}
\item
\[
\int_\Theta{\left|\overline{\pi}_{(m,n)}(\theta)-\pi_{(n)}(\theta)\right|}d\theta =O\left(\frac{1}{nm^{2}}\right).
\]
\item Suppose that $h(\theta)$ is a function such that $\lim \sup \E_{\pi_{(n)}}[h^2(\theta)]<\infty$.
Then
\[
\left|{\E_{\overline{\pi}_{(m,n)}}[h(\theta)]-\E_{\pi_{(n)}}[h(\theta)]}\right|= O\left(\frac{1}{nm^{2}}\right).
\]
\end{enumerate}
\end{thm}
The proof is in Section~\ref{AppendixTop:Proofs}.

Note first that for a fixed $n$ the errors in Theorem~\ref{thm:maintheorem} are of order $O(m^{-2})$ in the subsample size. More importantly, the theorem shows that the perturbation error can decrease at a very rapid rate with respect to $n$. For example, $m = O(n^\frac12)$ gives a perturbation error of order $O(n^{-2})$. However, the accuracy of the control variates expanded around the posterior mode increases so extremely rapidly with the sample size $n$ that the optimal subsample size $m = O(n^{-1})$ actually decreases with $n$. This in turn leads to an perturbation error of $O(n)$. Control variates based on expanding around the posterior mode therefore makes the two aims efficiency and accuracy incompatible.

However, it is not practical to use control variates based on the posterior mode as we wish to avoid handling all the observations. A way around this is to obtain the posterior mode using Stochastic Gradient Decent (SGD) based on unbiased estimate of the gradient from a subsample. Alternatively, one can use the posterior mode from a fixed subsample. The following corollary shows the approximation rates in Theorem~\ref{thm:maintheorem} and the asymptotic behavior of $\sigma_{LL,m,n}^{2}$ in Lemma \ref{lem:m_and_n_asymptotics} when the control variates are based on the posterior mode from a fixed subset of $\wt n \ll n$ observations. Its proof is in Section~\ref{AppendixTop:Proofs}.

\begin{corollary} \label{corr: cv implications} Suppose that $\theta_{\wt n}^\star -
\theta_n^\star = O(\wt n^{-\frac12})$ and Assumptions~\ref{Ass2} or \ref{ass: assump for mle}
hold. Then,
\begin{enumerate}
\item
\[
\int_\Theta{\left|\overline{\pi}_{(m,n)}(\theta)-\pi_{(n)}(\theta)\right|}d\theta =O\left(\frac{n^2}{m^{2}\wt n^3}\right).
\]
\item Suppose that $h(\theta)$ is a function such that $\lim \sup \E_{\pi_{(n)}}[h^2(\theta)]<\infty$.
Then
\[
\left|{\E_{\overline{\pi}_{(m,n)}}[h(\theta)]-\E_{\pi_{(n)}}[h(\theta)]}\right|= O\left(\frac{n^2}{m^{2}\wt n^3}\right).
\]
\item
$\sigma_{LL,m,n}^{2}(\theta)  = O\left(\frac{n^2}{m\wt n^3}\right)$ for $\Sigma_n^{-\frac12}(\theta-\theta_n^\star)=O(1)$.
\end{enumerate}
\end{corollary}
To understand the implications of this result, suppose that $\wt n = n^\kappa, m = n^\alpha $ and  we target
$\sigma_{LL,m,n}^{2}(\theta) = O(1)$. Then, Corollary~\ref{corr: cv implications} (iii) implies that the optimal subsample is obtained with $\alpha= 2 - 3\kappa$. The errors in (i) and (ii) then decrease with $n$ if only if $\kappa < 2/3$. If we for example take $\kappa = 1/2$, then $\alpha = 1/2 $ and the error in
parts (i) and (ii) of  Corollary~\ref{corr: cv implications} are $O( n^{-\frac12})$. If instead $\kappa \geq 2/3$ then $\alpha \leq 0$, so the optimal $m$ is decreasing in $n$, and the errors in parts (i) and (ii) therefore increase with $n$. So for $\kappa \geq 2/3$ there is a tradeoff between efficiency and accuracy.

An interesting intermediate approach uses $\wt n \ll n$ observations for the control variates initially and then updates $\theta_{\wt n}^\star$ after the sampler has reached a central region in the posterior. This would correspond to using a $\kappa$ closer to one, with the approximation error rates being closer to those in Theorem \ref{thm:maintheorem}.

Finally, we note that it is straightforward to show that Theorem 1 still holds if we construct the control variates using the MLE rather than a posterior mode. To do so we assume that

\begin{ass}\label{ass: assump for mle} In Assumption~\ref{Ass2} we replace $\pi_{(n)}(\theta)$ by $L_{(n)}(\theta)$, so that $\theta_n^\star$ is now an MLE,
$\Delta_n(\theta) = \partial \ell_{(n)} (\theta)/\partial \theta \partial {\theta^{T}}$, etc.
\end{ass}
\noindent Then Theorem~\ref{thm:maintheorem} holds under Assumption~\ref{ass: assump for mle} and mild conditions on the prior, e.g. that $p_\Theta(\theta)/p_\Theta(\theta_n^\star)$ is bounded.

\subsection{Approximating the perturbation error\label{subsec:A-heuristic-to}}

Theorem \ref{thm:maintheorem} and Corollary \ref{corr: cv implications} are large sample results on the error in the perturbed posterior. In this section we give sharper, but more heuristic, results on this propotional error in the perturbed posterior and show that it is a lot smaller that the proportional error in the perturbed likelihood. We then outline how these sharper bounds can be used to estimate the  proportional error in practice.

Let $\xi_{m,n}(\theta) = \wh \ell_{(m,n)} (\theta) - \frac12 \wh \sigma^2_{LL,m,n} (\theta)$. Then, we can show that
for large $m$, $\E(\xi_{m,n}(\theta) ) = \ell_{(n)}(\theta) - \frac12 \sigma^2_{LL,m,n} (\theta)$ and
$\Lambda_{(m,n)}(\theta) = \V ( \xi_{m,n}(\theta)) = \sigma^2_{LL,m,n} (\theta) + 2\Gamma_{(m,n)}(\theta)$, where
\begin{align} \label{eq: defn of Gamma_mn}
\Gamma_{(m,n)}(\theta)& =\frac{\sigma_{LL,m,n}^{4}(\theta)}{8m}\left(\Psi_{d,n}^{(4)}(\theta)-1\right)-\frac{\sigma_{LL,m,n}^{3}(\theta)}{2\sqrt{m}}\Psi_{d,n}^{(3)}(\theta).
\end{align}
where $\Psi_{d,n}^{(b)}\coloneqq\varphi_{d,n}^{(b)}/\sigma_{d,n}^{b}$
for $b=1, \dots, 4$.

We now take $m = m(n)$, e.g. $m = O(\sqrt n) $ and suppose that as $n \rightarrow \infty$, $ \sigma^2_{LL,m,n} (\theta)
\rightarrow \ov \sigma^2_{LL,m,n} (\theta)< \infty $ and $\Psi_{d,n}^{(b)}(\theta) \rightarrow \ov \Psi_{d,n}^{(b)}(\theta)$, with $\ov \Psi_{d,n}^{(b)}(\theta)$ bounded for all $\theta$. Then, by a standard central limit argument we can show that
$\xi_{m,n}(\theta) - \left ( \ell_{(n)}(\theta) - \frac12 \sigma^2_{LL,m,n} (\theta)\right)$ tends to a normal density
with mean 0 and variance $\ov \sigma^2_{LL,m,n} (\theta)$.

This central limit theorem result is driven by $m$ becoming large. Hence, if $n$ is fixed and $m \uparrow m(n)=\sqrt n $ we will have that
$\xi_{m,n}(\theta)- \ell_{(n)}(\theta) - \left ( \ell_{(n)}(\theta) - \frac12 \sigma^2_{LL,m,n} (\theta)\right) $ tends to a normal with variance
 $\Lambda_{(m,n)}(\theta)$. Now for fixed $n$, $\xi_{m,n}(\theta)- \ell_{(n)}(\theta) - \left ( \ell_{(n)}(\theta) - \frac12 \sigma^2_{LL,m,n} (\theta)\right) $ is bounded so that
 \begin{align}\label{eq: clt approx}
\E \left [ \exp\left (\xi_{m,n}(\theta)- \ell_{(n)}(\theta) - \left ( \ell_{(n)}(\theta) - \frac12 \sigma^2_{LL,m,n} (\theta)\right) \right)\right ] & \rightarrow \exp\left ( \frac12 \Lambda_{(m,n)} (\theta)   \right )
\end{align}

Lemma \ref{lem:PerturbedLogLikRatio} below gives analytical expression for the proportional errors in the
perturbed likelihood $L_{(m,n)}(\theta)$ and the perturbed posterior. Its proof is straightforward and omitted.
The normality assumption in the lemma assumes that $n$ and $m = m(n)$ are large and is based on
\eqref{eq: clt approx}.

\begin{lemma}\label{lem:PerturbedLogLikRatio}
Suppose that
$\xi_{m,n}(\theta)$ is normal with mean  $\ell_{(n)}(\theta) - \frac12 \sigma^2_{LL,m,n} (\theta)$ and
variance $\Lambda_{(m,n)}(\theta)$
Then,
\begin{align}
\frac{L_{(m,n)}(\theta)- L_{(n)}(\theta)}{L_{(n)}(\theta) }
& =  \exp\left (\Gamma_{(m,n)}(\theta)\right)-1 ,\label{eq:PerturbedLikelihood}
\end{align}
is the proportional error in the perturbed likelihood and
\begin{align} \label{eq: perturb post}
\frac{ \pi_{(m,n)} (\theta) - \pi_{(n)} (\theta)}{ \pi_{(n)} (\theta) }  & =  \frac{  \exp\left (\Gamma_{(m,n)} (\theta )\right ) }{ \E_{\pi_{(n)}} \left (  \exp\left (\Gamma_{(m,n)}\right ) \right)}
-1
\end{align}
is the proportional error in the perturbed posterior.
\end{lemma}
From part (iii) of Lemma~\ref{lem:m_and_n_asymptotics}, $\Psi_{d,n}^{(b)}(\theta)=O(1)$ for any $b\geq1.$ Hence, it follows from Lemma~\ref{lem:PerturbedLogLikRatio} that the perturbation
error \eqref{eq:PerturbedLikelihood} in the likelihood  depends
on $\sigma^2_{LL,m,n}(\theta)$ whereas the error in the perturbed posterior
\eqref{eq: perturb post}
 will tend to be much smaller because the term
\begin{align*}
\frac{  \exp\left (\Gamma_{(m,n)} (\theta )\right ) }{ \E_{\pi_{(n)}} \left (  \exp\left (\Gamma_{(m,n)}\right ) \right)}
\end{align*}
will be close to 1 for all $\theta $ in the region $||\Sigma_n^{-\frac12} (\theta - \theta_n^\star)||\le k $ for a fixed $k>0$ as the posterior becomes very concentrated around $\theta_n^\star$ for $n$ large. In particular,
if we write $\Gamma_{(m,n)} (\theta ) = C + \gamma_{(m,n)} (\theta)$ where $C$ is independent of $\theta$
and suppose that $\gamma_{(m,n)} (\theta)\ll C$. Then, the proportional error in the perturbed likelihood depends on $C$, whereas the error in the perturbed posterior
\begin{align*}
\frac{  \exp\left (\Gamma_{(m,n)} (\theta )\right ) }{ \E_{\pi_{(n)}} \left (  \exp\left (\Gamma_{(m,n)}\right ) \right)}& = \frac{  \exp\left (\gamma_{(m,n)} (\theta )\right ) }{ \E_{\pi_{(n)}} \left (  \exp\left (\gamma_{(m,n)}\right ) \right)}
\end{align*}
will be very small. If $\gamma_{(m,n)} (\theta )\equiv 0 $, then there is no approximation in the perturbed posterior
even if $C$ is large so that the error in the perturbed likelihood is large. Thus, the error in the perturbed posterior is likely to be much smaller than in the perturbed likelihood.

We can use Lemma~\ref{lem:PerturbedLogLikRatio}
to estimate the perturbation error in the posterior for any given application.
The term $\Gamma_{(m,n)}(\theta) $ can be evaluated or estimated from a subsample because
the  terms $\sigma_{LL,m,n}(\theta)$
and $\Psi_{d,n}^{(b)}(\theta)$ are easily evaluated for any $\theta$
at the cost of evaluating $\ell_{i}(\theta)$ for all $i=1,...,n$,
or estimated from a subsample. The term
$\E_{\pi_{(n)}} \left (  \exp\left (\Gamma_{(m,n)}\right ) \right)$ can be estimated from the MCMC output.
Alternatively, we can use a Laplace approximation by
taking $\pi_{(n)} (\theta)$ as approximately normal with mean $\theta_n^\star$ and covariance matrix
$\Sigma_n$ and then approximate $\Gamma_{(m,n)}(\theta)$ by a quadratic centered at $\theta_n^\star$, where
$\theta_n^\star$ is obtained from the MCMC output.

\begin{rem}
Similar results to the above can be obtained if $\sigma^2_{LL,m,n} (\theta )/m^\beta \rightarrow \ov \sigma^2_{LL,m,n}(\theta )$ as $n \rightarrow \infty$, with $0<\beta< 1 $.
\end{rem}

\subsection{Subsampling with correlated proposals of $u$\label{subsec:Correlated-updates-of-u}}
\citet{deligiannidis2015correlated}
propose a general method that correlates the current and proposed
values of the $u_{i}$. The advantage of using this correlation is that it makes
the variance of the difference in the logarithms of the estimated
likelihoods in \eqref{eq:accprobPseudoMCMC} much smaller
than that of each of the terms themselves. This leads, in our context,
to being able to target much higher values of $\sigma_{LL,m,n}^{2}(\theta)$ than unity
thus requiring much smaller values of $m$. In this section, we adapt the method of
\citet{deligiannidis2015correlated} to our problem, and in the next
we discuss a variant of the correlated pseudo-marginal which we call the block correlated
pseudo marginal.

For the correlated PM approach to subsampling, we let $u$ be a vector
of length $n$ with binary elements $u_{i}$ that determine if observation
$i$ is included ($u_{i}=1$) when estimating the log-likelihood.
Note that this is different from above, where $u$ contained the
observation indices and was of length $m$. Moreover, here the sample
size is random and we let $m^{\star}$ be the expected sample size.
The sampling probabilities become $\Pr(u_{i}=1)=m^{\star}/n$ for
$i=1,\dots,n$. We use the auxiliary variable (particle) $v$ in \citet{deligiannidis2015correlated}
to induce dependence at the current $u_{i}^{c}$ and proposed $u_{i}^{p}$
sampling indicator through a Gaussian copula as we now explain. The
correlated pseudo-marginal method uses a Gaussian autoregressive
kernel $\mathcal{K}(v_{c},v_{p})$ defined by $v_{p}=\phi v_{c}+\sqrt{1-\phi^{2}}\varepsilon$,
where $\varepsilon\sim\mathcal{N}(0,1)$. We also have $v_{c}\sim p(v)=\mathcal{N}(v|0,1)$
and $\mathcal{K}(v_{c},v_{p})$ is reversible with respect to $p(v)$.
We sample the $u_{i}$'s by first generating $v_{c}$ and $v_{p}$
and set $u_{i}^{c}=\mathcal{I}\left[\Phi(v_{c})\leq\frac{m^{\star}}{n}\right]$
and $u_{i}^{p}=\mathcal{I}\left[\Phi(v_{p})\leq\frac{m^{\star}}{n}\right]$,
where $\Phi$ denotes the standard normal cdf.

As noted above, in contrast to Section~\ref{subsec:DE}, $u$ is a
binary vector. We can instead use the Horvitz-Thompson \citep{horvitz1952generalization}
which (under SRS) is
\[
\widehat{d}_{(m^{\star},n)}=\mathlarger{\sum}_{i=1}^{n}\frac{d_{i,n}}{m^{\star}/n}u_{i},
\]
and is unbiased for $d_{(n)}$. Note that we can write
\[
\widehat{d}_{(m^{\star},n)}=\frac{1}{m^{\star}}\sum_{i=1}^{n}nd_{i,n}u_{i},\quad\text{with }\sigma_{LL,m^{\star},n}^{2}=\frac{\sigma_{\xi,m^{\star},n}^{2}}{m^{\star}},\quad\text{where }\sigma_{\xi,m^{\star},n}^{2}=n\left(1-\frac{m^{\star}}{n}\right)\sum_{i=1}^{n}(d_{i,n}-\mu_{d,n})^{2}
\]
can be unbiasedly estimated by
\[
\widehat{\sigma}_{\xi,m^{\star},n}^{2}=n^{2}\left(1-\frac{m^{\star}}{n}\right)\frac{1}{m^{\star}}\sum_{i=1}^{n}(d_{i,n}-\mu_{d,n})^{2}u_{i}.
\]

\subsection{Subsampling with block proposals for $u$}\label{SS: block proposals}

\citet{tran2016block} propose the block correlated PM algorithm and show that
it is a natural way to correlate the logs of the likelihood estimates at
the current and proposed value of the parameters in our subsampling problem. The method divides
the vector of observation indices $u=(u_{1},\dots,u_{m})$ into $G$
blocks and then updates one block at a time jointly with $\theta$.
By setting a large $G$, a high correlation $\rho$ between the estimates of the likelihoods
at the proposed and current parameter values is induced, reducing
the variability of the difference in the logs of the estimated likelihoods at the proposed and current values of
$\theta$. More precisely, they show
that under certain assumptions $\rho$ is close to $1-1/G$.

\subsection{Optimal variance of the estimator\label{subsec:Choosing-the-sampling}}

\citet{pitt2012some}, \citet{doucet2012efficient} and \citet{sherlock2013efficiency}
obtain the value of $\V(\log \widehat L)$, where $\widehat L$ is an unbiased likelihood estimator (e.g. based on importance sampling or a particle filter) that optimizes the trade off between MCMC sampling efficiency and computational cost in standard
PM. The consensus is that $\V(\log \widehat L)$ should lie in the interval
$[1,3.3]$, where the less efficient the proposal in
the exact likelihood setting, the higher the optimal value of $\V(\log \widehat L)$.
The optimal value is derived assuming that the cost of computing one
MCMC sample is inversely proportional to $\V(\log \widehat L)$.

For our problem, the log of the estimated likelihood is
$\log \left ( \wh L_{(m,n)}(\theta)\right ) = \wh \ell_{(m,n)}(\theta) - \frac12 \wh \sigma^2_{LL,m,n} (\theta)$,
which has variance $\Lambda_{(m,n)}(\theta)=\sigma^2_{LL,m,n}(\theta) + 2\Gamma_{(m,n)}(\theta)$,
where $\Gamma_{(m,n)}(\theta)$ is defined in \eqref{eq: defn of Gamma_mn}.
We take the computing cost as inversely proportional to $\sigma^2_{LL,m,n}(\theta)$
because our estimation effort is based on computing $\wh \ell_{(m,n)}$, with the extra cost of computing
$\wh \sigma^2_{LL,m,n}$ being negligible in comparison.

Thus,  for the parameter expanded control variates
we follow \cite{pitt2012some} and define the computational time as
\begin{align}
\mathrm{CT}(\sigma_{LL,m,n}^{2},\Lambda_{(m,n)}) &:=  \frac{\mathrm{IF}(\Lambda_{(m,n)})}{\sigma_{LL,m,n}^{2}},
\text{ with }\text{ }\mathrm{IF}(\Lambda_{(m,n)}):=1+2\sum_{l=1}^{\infty}\rho_{l}, \label{eq:CT_invpropto}
\end{align}
which is proportional to the time required to produce one sample equivalent to an i.i.d. draw from the posterior distribution.
In~\eqref{eq:CT_invpropto},  $\rho_{l}$ is the $l$-lag autocorrelation of the chain and
$\mathrm{IF}(\Lambda_{(m,n)}) $ is the Inefficiency Factor (IF), which we note depends on $\Lambda_{(m,n)}$. However,
$\Lambda_{(m,n)} \approx \sigma^2_{LL,m,n}$ for $m$ large so that we will write $\mathrm{IF}(\sigma^2_{LL,m,n})$.

If we use the data expanded control variates, then it is necessary to select both $m$ and the number of clusters $K$.
The computational cost of each cluster involves computing $\ell_i, $ and its gradient and Hessian at the centroid. An approximate upper bound for the cost
of a new cluster is therefore $3c_{\ell}$, where $c_{\ell}$ is the
cost of a single $\ell_{i}$-evaluation. However, in many models it is possible to reuse
some terms  when computing the gradient and Hessian, so the
true cost is probably much closer to $c_{\ell}$. For example, in the logistic regression model in
Section~\ref{sec:Application}, the gradient and Hessian will be functions of $\exp(\pm x^T_i \theta)$ which is already computed when evaluating $\ell_i(\theta)$.
Assuming that the cost of a cluster is $\omega c_{\ell}$, for some $\omega>0$,
a reasonable measure of computational time is
\begin{align}
\mathrm{CT}_{(m,K)}(\sigma_{LL,m,n}^{2}(K)):=\mathrm{IF}(\Lambda_{(m,n)}^{2}(K))\times(\omega K+m).\label{eq:CT_proxies}
\end{align}
This expression is similar to \citet{tran2014importance} who also
take into account an overhead cost in their computational time. We find $m$ and $K$
by standard numerical optimization using an expression for the inefficiency ($\mathrm{IF}$)
(e.g. the ones derived in \citealp{pitt2012some} for PM and \citealp{tran2016block}
for block PM).

For the correlated PM, we can follow
\citet{deligiannidis2015correlated} and show for our application that the  variance of the log of the estimated
likelihood at the proposed values of $u$ and $\theta$ conditional on the
the estimated
likelihood at the current values of $u$ and $\theta$ is $\tau_{m,n}^{2}=\Lambda_{(m,n)} (1-\rho^{2})\approx
\sigma_{LL,m,n}^{2}(1-\rho^{2})$,
where $\rho$ is the correlation between the logs of the two estimates of the likelihood,
with the optimal value of $\tau_{m,n}^{2}$ around 4.
Similarly, for the
block correlated PM, \citet{tran2016block} show that the  variance of the log of the likelihood
estimator at the proposed values conditional on only updating one block of $u$, keeping
the others fixed, is $\tau_{m,n,G}^{2}=\Lambda_{(m,n)}(1-\rho_{G}^{2})\approx  \sigma_{LL,m,n}^{2}(1-\rho_{G}^{2})$.
Let $G=G(m)=O(m^{\beta})$. Using Corollary~\ref{corr: cv implications}
and $\rho_{G}(m)=1-1/G(m)$, it follows using the same notation as in the discussion
below that corollary that
$\tau_{m,n,G}^{2}(\theta) = O(1)$ is achieved if we take $m=O(n^{\alpha}), \wt n = n^{\kappa}$ with
$2 = 3\kappa + \alpha(1+\beta)$. If $\kappa = 1/2$ and $\beta = 0 $, i.e. $G$ does not depend on $m$,
then the approximations in parts (i) and (ii) of Corollary \ref{corr: cv implications}
are $O(n^{-\frac12})$. We can then ensure that $\tau_{m,n,G}^{2}(\theta) $ is around the optimal value of 4
while
$\sigma^2_{LL,m,n}\gg 1$ by adapting $G$.  In practice, we usually take $G=100$ which gives us a correlation close
to 0.99.

We emphasize that
it is the {\em combined} effect of using both the control variates and correlating the logs of the estimated
likelihoods at the current and proposed values
that makes the method scale well.

\subsection{Strategy for subsampling MCMC\label{subsec:Strategy-for-subsampling}}

We have argued that the parameter expanded control variates have good
asymptotic properties and that the data expanded control variates
have the advantage of not requiring a central measure $\theta^{\star}_n$
of $\theta$. Data expanded control variates also have the advantage of working well over the whole parameter space since they are always evaluated at the proposed $\theta$. Our proposed subsampling MCMC algorithm will therefore
begin with the data expanded control variates during a training period
and then switch to the parameter expanded control variates once
we have learned a reasonable value of $\theta^{\star}_n$. This value
is set at the end of the training period by computing the geometric
median \citep{vardi2000multivariate} of the $10$\% preceding iterations,
which requires evaluating the likelihood over the full dataset once. We include
this in our computational cost.

Although we have argued that the data expanded control variates have
poor asymptotic properties for large $p$, we can still use them with a reasonably
small $K$ as the error decreases at the fast rate $O(m^{-2})$. Hence, there is no need to make the approximation
very accurate by using a large $K$ in relation to $n$, as this
increases the computing cost.

\section{Applications\label{sec:Application}}

\subsection{Empirical studies}\label{SS: emp studies}
This section performs a number of experiments to compare our proposed algorithms against both standard MCMC which we call MH and other competing subsampling methods. To compare against other subsampling approaches we follow  \citet{bardenet2015markov}. We compare the standard (independent) PM, the correlated PM and block correlated PM and the with correlated PM subsampling using the data expanded control variates, since, for our examples,
the parameter expanded control variates will give a very small variance once we find a good $\theta^{\star}_n$, and hence there are no gains from implementing BPM or CPM compared to PM. However, note that correlating or blocking subsamples is especially
useful in the training phase of our algorithm that combines both types of control
variates as described in Section~\ref{subsec:Strategy-for-subsampling}, when we are learning
 about an appropriate $\theta^{\star}_n$, because otherwise the algorithm is likely to get stuck.
\subsection{Models and data sets\label{subsec:Models-and-data}}
We consider three models in our experiments. The first two, which are used for comparing against other subsampling approaches, are $\mathrm{AR}(1)$ models with Student-t iid errors $\epsilon_{t}\sim t(5)$
with $5$ degrees of freedom
\begin{align*}
\mathrm{M_{1}:\;} & y_{t}=\beta_{0}+\beta_{1}y_{t-1}+\epsilon_{t}\quad\quad\;\;\;\left[\theta=(\beta_{0}=0.3,\beta_{1}=0.6)\right]\\
\mathrm{M_{2}:\;} & y_{t}=\mu+\varrho(y_{t-1}-\mu)+\epsilon_{t}\quad\left[\theta=(\mu=0.3,\varrho=0.99)\right]
\end{align*}
with priors
\[
p(\beta_{0},\beta_{1})\stackrel{\small{\text{ind.}}}{=}\mathcal{U}(\beta_{0}|-5,5)\cdot\mathcal{U}(\beta_{1}|0,1)\quad\text{and}\quad p(\mu,\varrho)\stackrel{\small{\text{ind.}}}{=}\mathcal{U}(\mu|-5,5)\cdot\mathcal{U}(\varrho|0,1),
\]
where $\mathcal{U}(\cdot|a,b)$ is the uniform density on the interval
$[a,b]$. Model $\text{M}_{2}$, the so called steady state AR, is
interesting as $\varrho$ close to $1$ gives a weakly identified
$\mu$, with a posterior that concentrates very slowly as $n$ increases
\citep{VillaniJAE2009}. We simulate $n=100,000$ observations from
both models.

The third model is the logistic regression
\begin{eqnarray*}
p(y_{i}|x_{i},\beta) & = & \left(\frac{1}{1+\exp(x_{i}^{T}\beta)}\right)^{y_{i}}\left(\frac{1}{1+\exp(-x_{i}^{T}\beta)}\right)^{1-y_{i}},\text{ with }\quad p(\beta)=\mathcal{N}(\beta|0,10I),
\end{eqnarray*}
which we fit to three datasets. The first dataset concerns
firm bankruptcy with $n=4,748,089$ observations with firm
default as the response variable and eight firm-specific and macroeconomic
covariates ($p=9$ with intercept); see \citet{giordani2011taking}
for details. We use this data set to study the different
proposals for $u$ with two proposals for $\theta$, the random walk MH and the independence
MH. The second
dataset is the well known HIGGS data \citep{baldi2014searching} with the response `detected
particle' explained by 21 covariates consisting of kinematic properties
measured by particle detectors (we exclude high-level features
for simplicity). From the 11 million observations we use a subset
of $n=$1,100,000 observations. The third dataset is Cover Type (Covtype)
which was originally  a classification problem with 7 classes.
We follow \citet{collobert2002parallel} and \cite{bardenet2015markov}
and transform it into a binary classification problem. The dataset consists of $n=550,087$
observations and $p=11$ variables, after removing the qualitative
variables for simplicity.
 We use these three datasets
 to benchmark our proposed subsampling MCMC algorithm in
Section~\ref{subsec:Strategy-for-subsampling} against standard MCMC using
a random walk MH proposal.

\subsection{Experiment 1: Comparing different proposals for $u$\label{subsec:Experiment-1}}

The first comparison between the different proposals for $u$ uses
the logistic regression with the Bankruptcy dataset described in
Section~\ref{subsec:Models-and-data}. Since there are relatively  few
observations corresponding to bankruptcy
$(y_{i}=1)$  ($41,566$ defaults), we only subsample the
observations with $y_{k}=0$, i.e., the first term in
\begin{eqnarray*}
\ell(\theta) & = & \sum_{i:y_{i}=1}\ell_{i}(\theta)+\sum_{i:y_{i}=0}\ell_{i}(\theta),
\end{eqnarray*}
is always evaluated (and included in the computational cost, $\mathrm{CC}$).

The tuning parameters $m$ and $K$ are determined by optimizing the
computational time $\mathrm{CT}$ in \eqref{eq:CT_proxies} with respect to $m$ and $K$, with
\[
\sigma_{LL,m,n}^{2}(K)=\frac{n^{2}\sigma_{d,n}^{2}(K)}{m}.
\]
We estimate the relation $\sigma_{d,n}^{2}(K)=C_{0}K^{\nu}$
 for each example by  running the clustering algorithm on a grid of $K$
and for each value of the grid we compute $\sigma_{d,n}^{2}$ at the
maximum likelihood estimator $\theta^{\star}_n$. Given $C_{0}$ and
$\nu$, it is straightforward to use the expression for the IF in
\citet{pitt2012some} (PM) and \citet{tran2016block} (block PM) to
minimize $\mathrm{CT}_{(m,K)}$ in \eqref{eq:CT_proxies} and obtain
$m_{\mathrm{opt}}$ and $K_{\mathrm{opt}}$ and the corresponding
$\sigma_{\mathrm{opt}}^{2}=\sigma_{LL,m_{\mathrm{opt}},n}^{2}(K_{\text{\ensuremath{\mathrm{opt}}}})$.
The correlated PM uses $m_{\mathrm{opt}}^{\star}=m_{\mathrm{opt}}$
and the same value of $K_{\mathrm{opt}}$ as the block correlated PM. Table \ref{tab:SettingsPMCMC}
summarizes the settings for comparing the proposals for $u$,
including the settings for the AR example in Section~\ref{subsec:Experiment-2}. Finally, we set $G=100$ ($\rho_G=0.99$) for the block PM
and $\phi=0.9999$
($\kappa=0.9863$) for the correlated PM.
\begin{table}[htbp]
\centering \caption{\emph{Experimental settings for comparing proposals for $u$ in the applications}.
$n$ is the number of observation. The proposals are the Random Walk
Metropolis (RWM) $q(\theta|\theta_{c})=\mathcal{N}(\theta|\theta_{c},\Sigma_{\theta^{\star}_n})$
and the Independent MH (IMH) $q(\theta)=t_{10}(\theta|\theta^{\star}_n,\Sigma_{\theta^{\star}_n})$,
where the location parameter is $\theta^{\star}_n$ is the posterior
mode and $\Sigma_{\theta^{\star}_n}$ is the negative inverse Hessian
of the log-posterior evaluated at $\theta^{\star}_n$, both obtained
from an initial numerical optimization. We denote the optimal sample
size and number of clusters by $m_{\mathrm{opt}}$ and $K_{\mathrm{opt}}$,
and $\sigma_{LL,\text{opt}}^{2}$ is the corresponding optimal variance
of the log-likelihood estimate. We use $N=50,000$ iterates after
discarding 5,000 iterates as burn-in.}

\begin{tabular}{llcccccccccc}
\toprule
 & \textbf{\footnotesize{}Example} &  & {\footnotesize{}$n$} &  & \textbf{\footnotesize{}Proposal} &  & {\footnotesize{}$100m_{\mathrm{opt}}/n$} &  & {\footnotesize{}$100K_{\mathrm{opt}}/n$ } &  & {\footnotesize{}$\sigma_{LL,\mathrm{opt}}^{2}$}\tabularnewline
\cmidrule{2-2} \cmidrule{4-4} \cmidrule{6-6} \cmidrule{8-8} \cmidrule{10-10} \cmidrule{12-12}
 & \textbf{\footnotesize{}Logistic} &  & {\footnotesize{}4.7$\times10^{6}$ } &  & {\footnotesize{}RWM/IMH } &  &  &  &  &  & \tabularnewline
\cmidrule{2-2}
 & {\footnotesize{}Uncorr} &  &  &  &  &  & {\footnotesize{}8.615} &  & {\footnotesize{}4.967} &  & {\footnotesize{}0.27}\tabularnewline
 & {\footnotesize{}Block / Corr} &  &  &  &  &  & {\footnotesize{}1.286} &  & {\footnotesize{}0.485} &  & {\footnotesize{}56.89}\tabularnewline
\cmidrule{2-2}
 & \textbf{\footnotesize{}AR(1): $\mathrm{M}_{1}$} &  & {\footnotesize{}$10^{5}$ } &  & {\footnotesize{}RWM } &  &  &  &  &  & \tabularnewline
\cmidrule{2-2}
 & {\footnotesize{}Uncorr} &  &  &  &  &  & {\footnotesize{}1.896 } &  & {\footnotesize{}2.464} &  & {\footnotesize{}0.11}\tabularnewline
 & {\footnotesize{}Block / Corr} &  &  &  &  &  & {\footnotesize{}0.757 } &  & {\footnotesize{}0.993} &  & {\footnotesize{}12.41}\tabularnewline
\cmidrule{2-2}
 & \textbf{\footnotesize{}AR(1): $\mathrm{M}_{2}$} &  & {\footnotesize{}$10^{5}$ } &  & {\footnotesize{}RWM } &  &  &  &  &  & \tabularnewline
\cmidrule{2-2}
 & {\footnotesize{}Uncorr} &  &  &  &  &  & {\footnotesize{}4.561 } &  & {\footnotesize{}8.192} &  & {\footnotesize{}0.11}\tabularnewline
 & {\footnotesize{}Block / Corr} &  &  &  &  &  & {\footnotesize{}2.151 } &  & {\footnotesize{}3.176} &  & {\footnotesize{}12.40}\tabularnewline
\bottomrule
\end{tabular}\label{tab:SettingsPMCMC}
\end{table}

\begin{figure}[h]
\includegraphics[width=1\columnwidth]{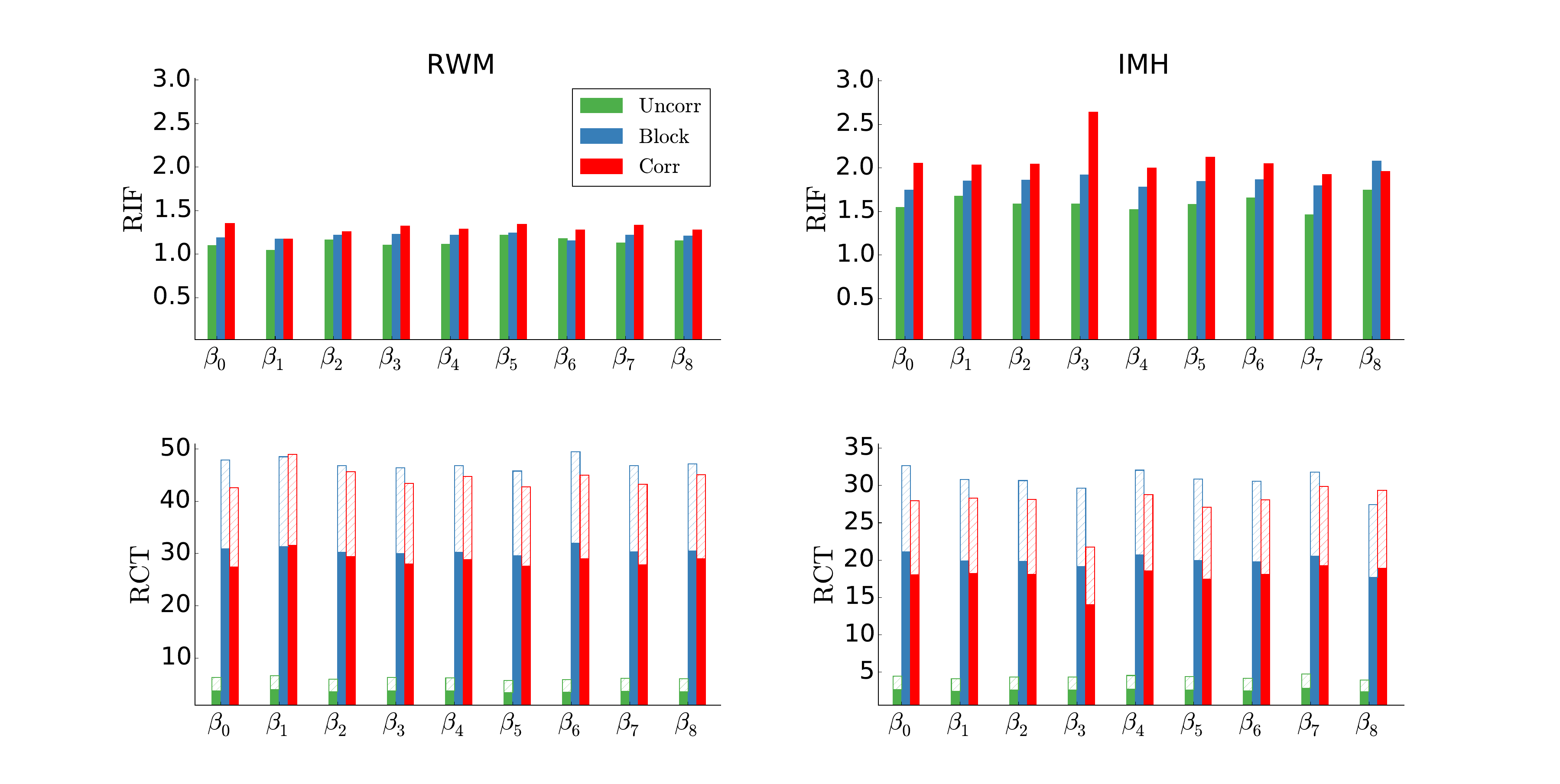}

\caption{\emph{Logistic regression for firm bankruptcy}. For algorithm $\mathcal{A}$ (uncorrelated
(Uncorr), block (Block) and correlated (Corr) PM) the figure shows
the Relative Inefficiency Factors (RIF) and Relative Computational
Time for RWM proposal (left panel) and IMH (right panel). For RCT,
the filled (dashed) bar correspond to $\omega=3$ ($\omega=1$) in
\eqref{eq:CT_proxies}.}
\label{fig:logistic_RIF_invRCT}
\end{figure}

\begin{figure}[h]
\includegraphics[width=1\columnwidth]{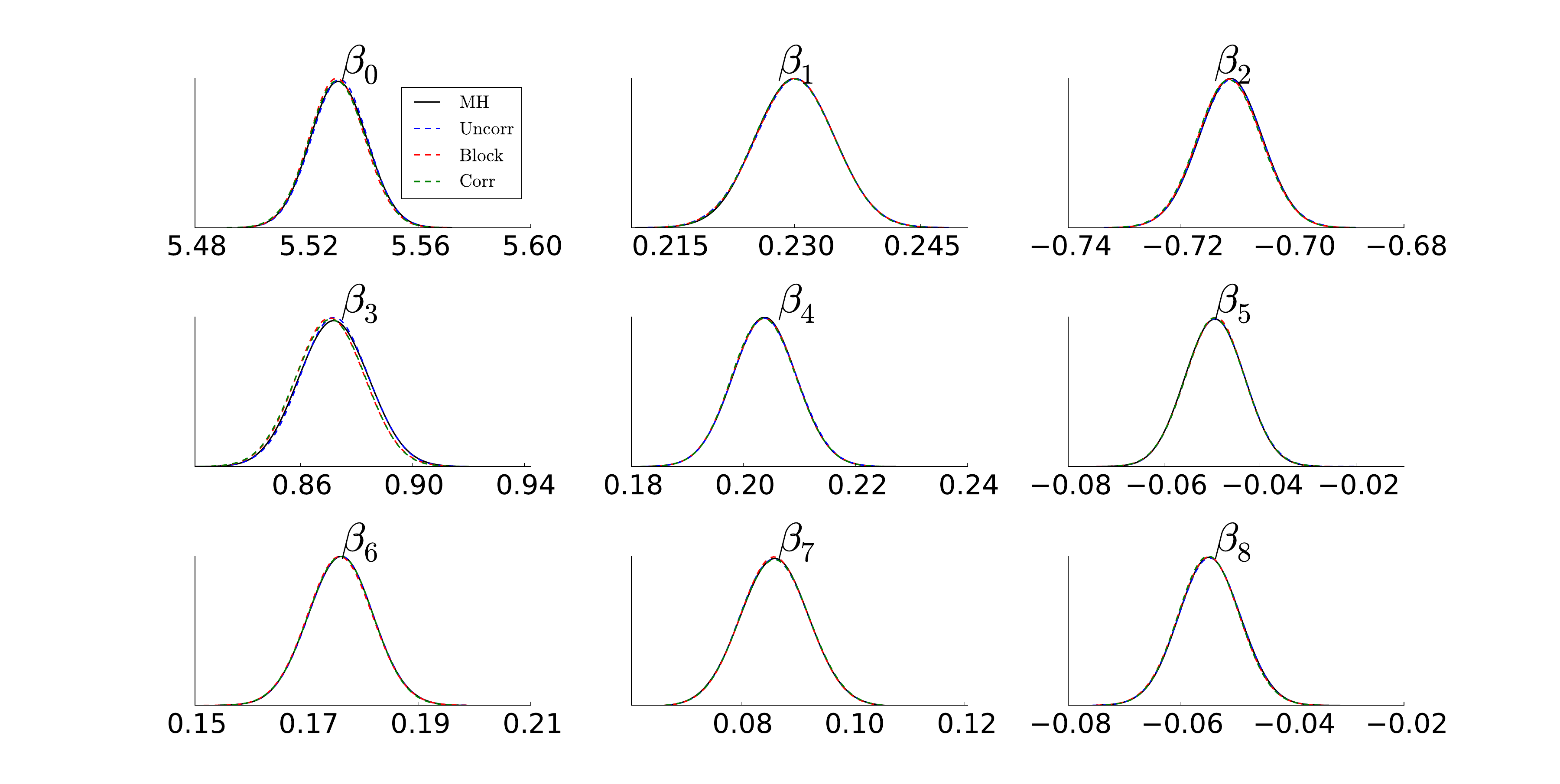}\caption{\emph{Logistic regression example}. Kernel density estimates of marginal
posteriors obtained by the IMH proposal. The figure shows the marginal
posteriors obtained using the uncorrelated (Uncorr), block (Block)
and correlated (Corr) PM (dashed blue, red and green, respectively)
and MH (solid black line).}
\label{fig:logistic_KDEs}
\end{figure}

Figure~\ref{fig:logistic_RIF_invRCT} shows the sampling efficiency
of the PM algorithms with the different proposals for $u$ relative
to that of the MH algorithm on the full dataset as measured by the Relative Computational
Time (RCT) defined, for any base sampler $\mathcal{A},$ as $\mathrm{CT}_{\mathrm{MH}}/\mathrm{CT}_{\mathcal{A}}$.
The figure also shows the Relative IF (RIF) , which is defined as
$\mathrm{IF}{}_{\mathcal{A}}/\mathrm{IF}_{\mathrm{MH}}$, where each
$\mathrm{IF}$ is estimated using the $\texttt{Coda}$ package in
$\texttt{R}$ \citep{coda2006}. The figure shows that both the correlated
and block PM schemes significantly outperform standard independent PM and also the MH algorithm applied to the full dataset with
respect to RCT. Figure~\ref{fig:logistic_KDEs} plots the Kernel Density
Estimates (KDE) of the posterior densities of the parameters for the three pseudo-marginal schemes
and the exact MH approach. The figure shows that targeting a large
$\sigma_{LL,m,n}^{2}$ ($\approx56$) for the block correlated and correlated
PM samplers results in a very small bias in this application, with
the proportional approximation error in  \eqref{eq: perturb post}
being $-0.01$ for both the block correlated and correlated PM and $-0.0001$
for the standard PM.


\subsection{Experiment 2: Comparison against other subsampling approaches\label{subsec:Experiment-2}}

We compare our algorithm against the approximate algorithms Austerity
MH \citep{korattikara2014austerity}, the confidence sampler \citep{bardenet2014towards},
the confidence sampler with control variates \citep{bardenet2015markov},
and the exact algorithm Firefly Monte Carlo \citep{maclaurin2014firefly}. See \citet{bardenet2015markov} for an excellent discussion of these algorithms.

We follow \citet{bardenet2015markov} in setting the tuning parameters of the competing algorithms, with the following exceptions. First,
we adapt during the burn-in phase to reach an acceptance probability
of $\alpha=0.35$ (instead of $\alpha=0.50$), which is optimal for
RWM with two parameters \citep{gelman1996efficient}. For the pseudo-marginals
we use $\alpha=0.15$ as in the five parameter example in \citet{sherlock2013efficiency}.
Second, the $p$-value of the $t$-test in the Austerity MH algorithm
is set to $\epsilon=0.01$ (instead of $\epsilon=0.05$) to put the
approximation error of the method on par with the other methods. Setting
$\epsilon=0.05$ gives an unusably poor approximation (and also produces
a much lower $\mathrm{RCT}$ than our methods). Additionally, the
confidence sampler with proxies (from a Taylor series approximation
with respect to $\theta$) requires that the third derivative can
be bounded uniformly for every observation and any $\theta$. This
bound is achieved by computing on a $\theta$-grid where the posterior
mass is located (this extra computational cost is not included in
the total cost here).

\begin{table}[htbp]
\centering \caption{\emph{AR-process example}. Mean of sampling fraction $f=m/n$ over
MCMC iterations for models $\mathrm{M}_{1}$ and $\mathrm{M}_{2}$
with MH (using the full data set), uncorrelated PM (Uncorr), block PM (Block) and correlated
PM (Corr), confidence sampler (Conf), confidence sampler with proxies
(ConfProxy), Austerity MH (AustMH), and Firefly Monte Carlo (Firefly).}

\begin{tabular}{rcccccccccccccccr}
\toprule
 & {\footnotesize{}$\mathrm{MH}$} &  & {\footnotesize{}$\mathrm{Uncorr}$} &  & {\footnotesize{}$\mathrm{Block}$} &  & {\footnotesize{}$\mathrm{Corr}$} &  & {\footnotesize{}$\mathrm{Conf}$} &  & {\footnotesize{}$\mathrm{ConfProxy}$} &  & {\footnotesize{}$\mathrm{AustMH}$} &  & {\footnotesize{}$\mathrm{Firefly}$} & \tabularnewline
\cmidrule{2-2} \cmidrule{4-4} \cmidrule{6-6} \cmidrule{8-8} \cmidrule{10-10} \cmidrule{12-12} \cmidrule{14-14} \cmidrule{16-16}
{\footnotesize{}$\mathrm{M}_{1}$} & {\footnotesize{}1.000 } &  & {\footnotesize{}0.093} &  & {\footnotesize{}0.037} &  & {\footnotesize{}0.037} &  & {\footnotesize{}1.493} &  & {\footnotesize{}0.160} &  & {\footnotesize{}1.037} &  & {\footnotesize{}0.100} & \tabularnewline
{\footnotesize{}$\mathrm{M}_{2}$} & {\footnotesize{}1.000 } &  & {\footnotesize{}0.291} &  & {\footnotesize{}0.117} &  & {\footnotesize{}0.116} &  & {\footnotesize{}1.490} &  & {\footnotesize{}1.500} &  & {\footnotesize{}1.019} &  & {\footnotesize{}0.137} & \tabularnewline
\bottomrule
\end{tabular}\label{tab:meanSamplingFrac}
\end{table}
\begin{figure}[h]
\includegraphics[width=1\columnwidth]{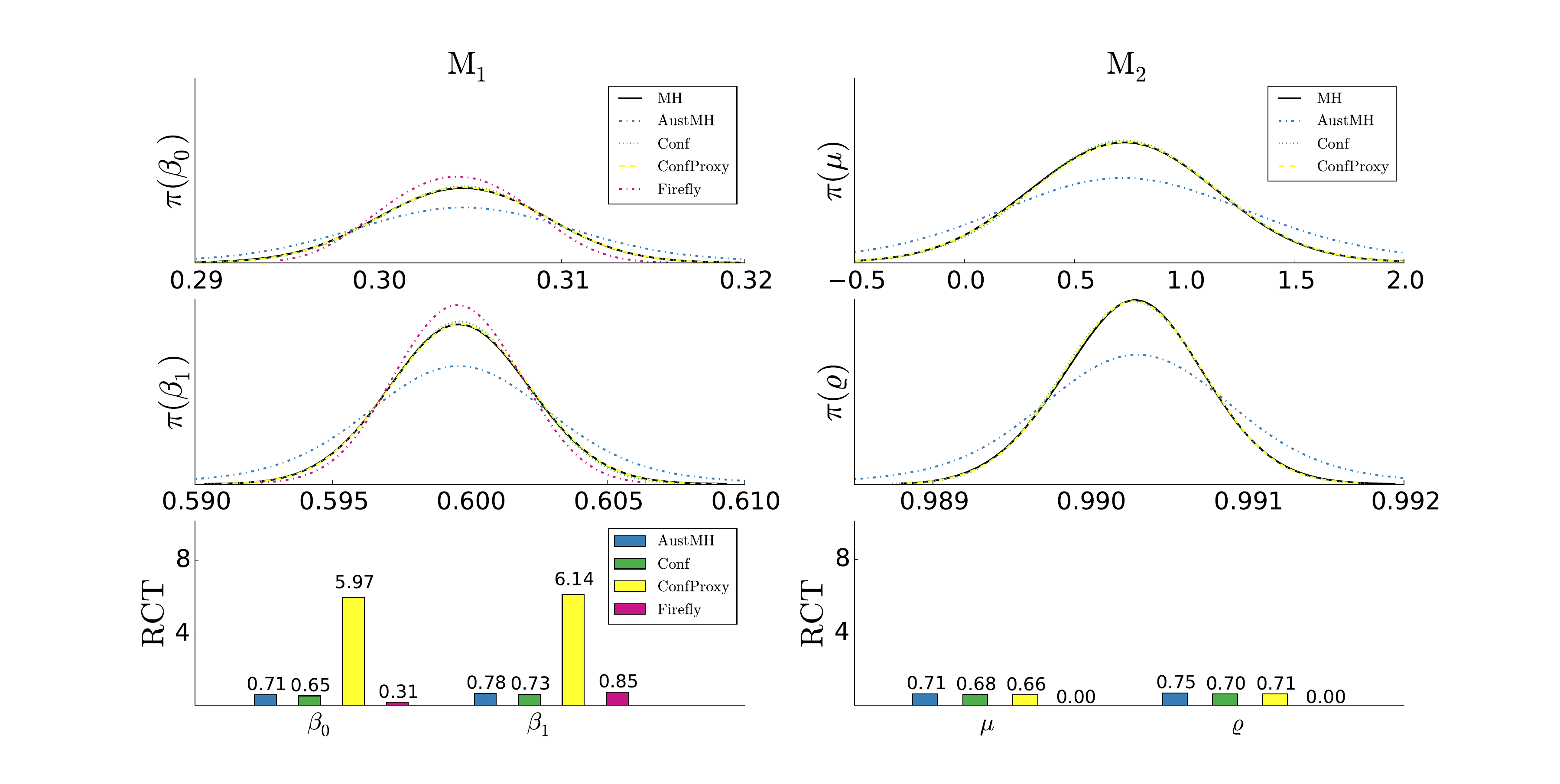}

\caption{\emph{AR-process example: Results for other subsampling algorithms}.
The left and right panel, respectively, show the results for model
$\mathrm{M}_{1}$ and $\mathrm{M}_{2}$. Each column shows the kernel
density estimates of marginal posteriors (top two) and for algorithm
$\mathcal{A}$ (confidence sampler (Conf), confidence sampler with
proxies (ConfProxy), Austerity MH (AustMH), and Firefly Monte Carlo
(Firefly)) the Relative Computational Time (RCT) (bottom).}
\label{fig:AR_KDE_invRCT-1_otherapproaches}
\end{figure}
\begin{figure}[h]
\includegraphics[width=1\columnwidth]{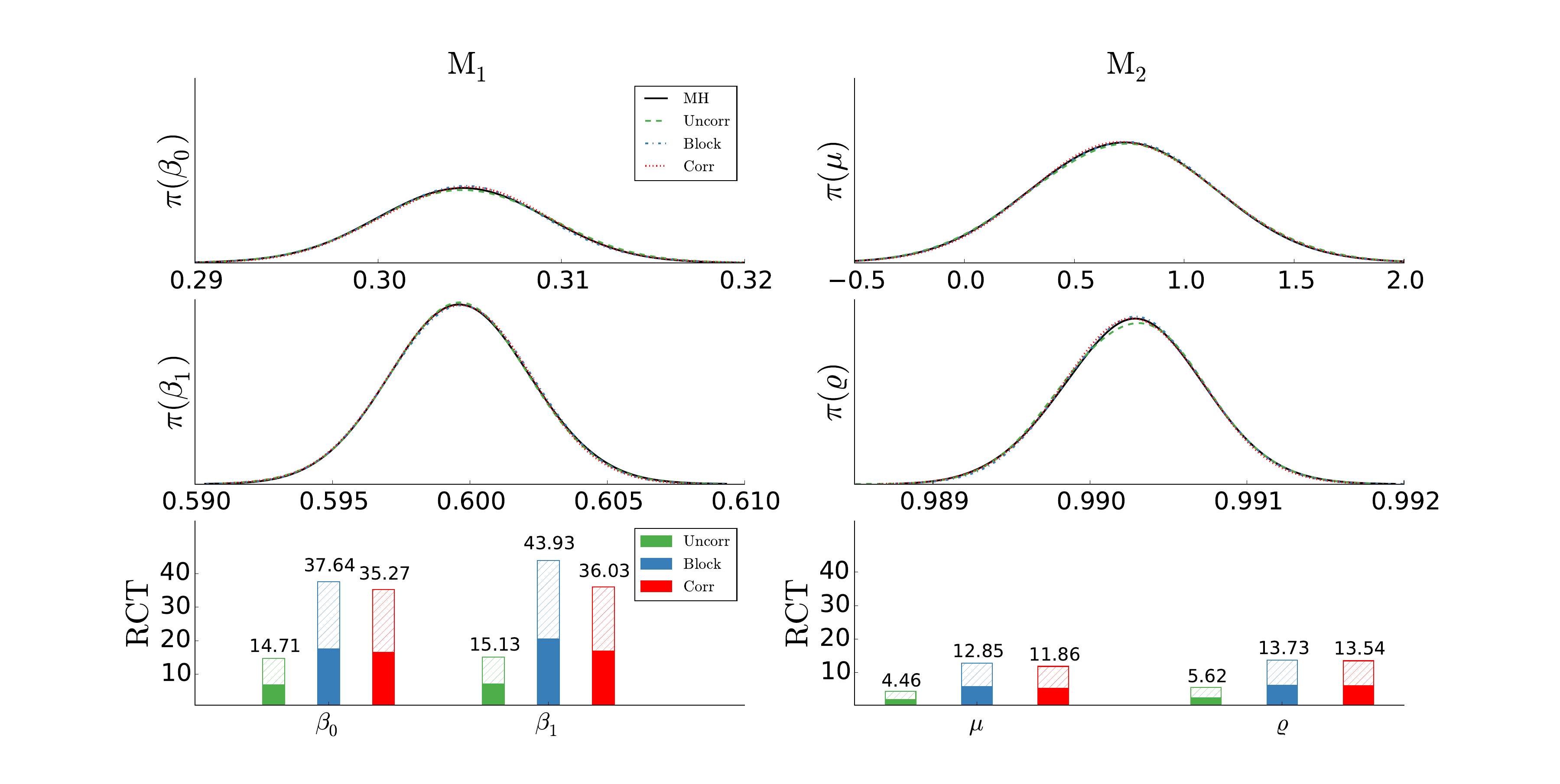}

\caption{\emph{AR-process example: Results for subsampling PM algorithms}.
The left and right panel, respectively, show the results for model
$\mathrm{M}_{1}$ and $\mathrm{M}_{2}$. Each column shows the kernel
density estimates of marginal posteriors (top two) and for algorithm
$\mathcal{A}$ (uncorrelated (Uncorr), block (Block) and correlated
(Corr) PM) the Relative Computational Time (RCT) (bottom). For RCT,
the filled (dashed) bar correspond to $\omega=3$ ($\omega=1$) in
\eqref{eq:CT_proxies}.}
\label{fig:AR_KDE_invRCT-1_PMMHapproaches}
\end{figure}
Table \ref{tab:meanSamplingFrac} shows the mean of the sampling fraction
over MCMC iterations. We note that both confidence samplers and the
Austerity MH estimate the numerator and denominator in each iteration,
and therefore require twice as many evaluations in a given iteration
as MCMC (in some cases evaluations from the previous iteration can
be reused). It is clear that our algorithms makes very efficient use
of a small subsample, especially the block and correlated PM samplers.

Figure~\ref{fig:AR_KDE_invRCT-1_otherapproaches} and \ref{fig:AR_KDE_invRCT-1_PMMHapproaches}
show the marginal posteriors obtained by, respectively, alternative
sampling approaches and the various PM approaches. Moreover, the figures
show the sampling efficiency of the different subsampling MCMC algorithms
relative to that of the MH algorithm as measured by the Relative Computational
Time. Figure~\ref{fig:AR_KDE_invRCT-1_otherapproaches} shows the
striking result that many of these approaches are not more efficient
than MH on the whole data set. The PM algorithms (and also the confidence samplers) provide excellent
approximations: indeed, the perturbation error in \eqref{eq: perturb post} is less than $10^{-6}$ for all our methods. Firefly Monte Carlo, although being an exact
algorithm, is highly inefficient in this example, as also documented
in \citet{bardenet2015markov}. In fact, for $\mathrm{M}_{2}$, we
were unable to obtain a single effective sample out of $55,000$
iterations, and hence it was impossible to construct a kernel density
estimate in this case.

We conclude that the only viable subsampling MCMC approaches are the
confidence sampler with proxies \citep{bardenet2015markov} and the
PM approaches we propose. Moreover, a significant speed up is only
obtained with the correlated PMs (both correlated and block).
\subsection{Experiment 3: Subsampling MCMC vs MCMC} \label{SS: subsampl MCMC vs MCMC}

Our final experiment compares standard MCMC against our algorithm
with a combination of control variates based on expanding $\theta$
and $z$ as described in Section~\ref{subsec:Strategy-for-subsampling}.
We use a random walk proposal with a scaled covariance matrix evaluated
at a $\theta^{\star}_n$ obtained from optimizing the posterior based
on $0.1$\% of the data. The same value is used as a starting value
for the algorithms. The scaling factor is $2.38/\sqrt{p}$ for MCMC
\citep{roberts1997weak} and $2.5/\sqrt{p}$ for subsampling MCMC
\citep{sherlock2013efficiency}. We set the training period (see
Section~\ref{subsec:Strategy-for-subsampling}) to $5000$ iterations and
sample $50,000$ draws thereafter. Our algorithm uses the  block PM for
updating $u$, where we set $m$ and $K$ following Section~\ref{subsec:Experiment-1}.
After the training period we reset $m$ as the initial $m$ is now
too large (since the control variates based on $\theta$ now give
an accurate approximation). We set the new value to $m=1,000$, which
is sensible for block PM with $G=100$.

\begin{figure}[h]
\includegraphics[width=0.75\columnwidth]{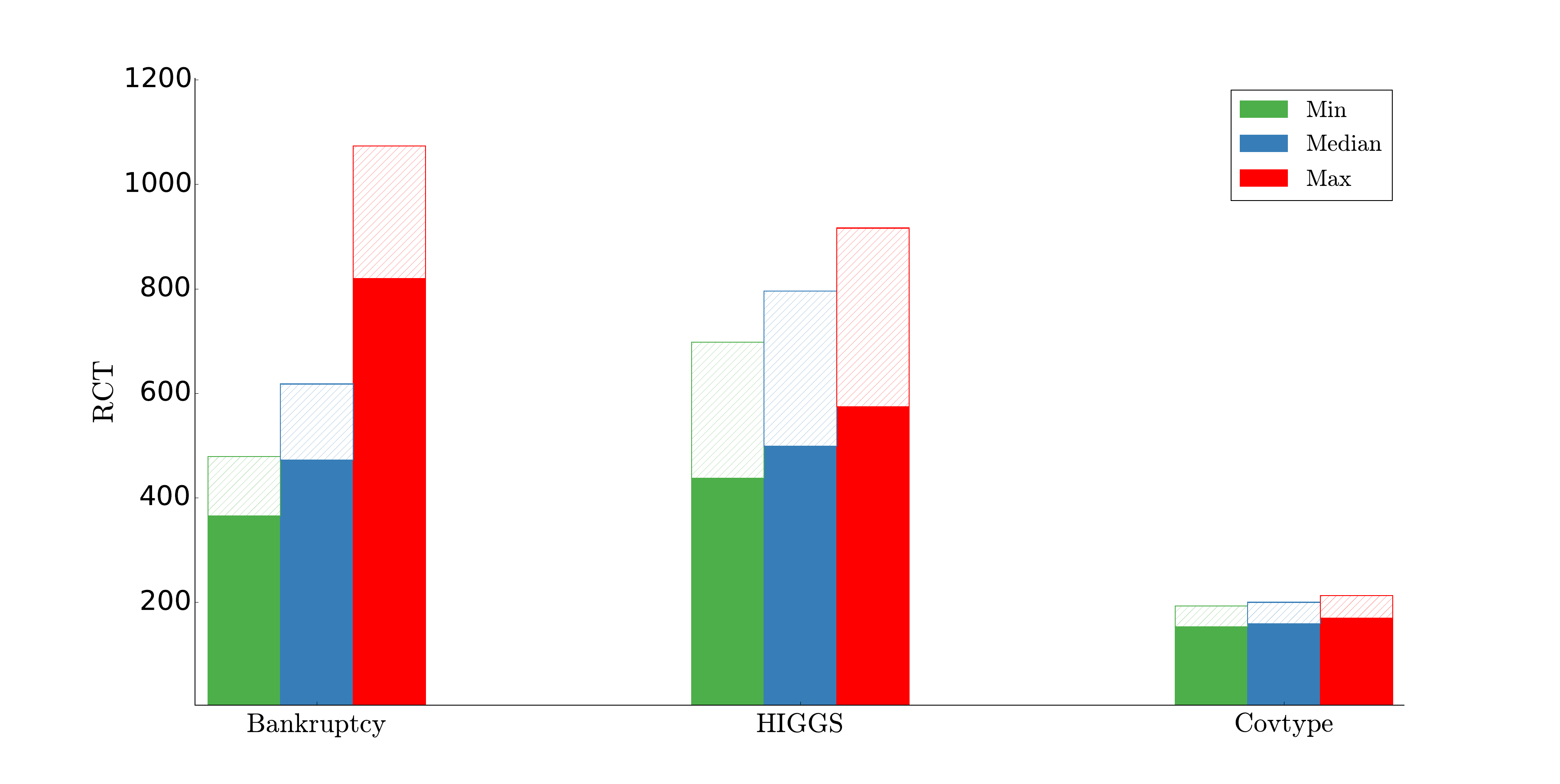}

\caption{\emph{Subsampling MCMC vs MCMC}. The figure shows Relative Computational
Time (RCT) for different data sets. The RCT over the parameters are
summarized by the minimum (green), median (blue) and maximum (red).
The PM algorithm combines the control variates based on expanding
$\theta$ and $z$ as described in Section~\ref{subsec:Strategy-for-subsampling}
and use block proposals for $u$. The filled (dashed) bars correspond
to the lower (upper) bound of the computational cost discussed
in Section~\ref{subsec:Computational-complexity}.}
\label{fig:RCT_3datasets}
\end{figure}

Figure~\ref{fig:RCT_3datasets} shows the $\mathrm{RCT}$ for each
of the data sets. Significant speed ups are achieved by switching
to the parameter expanded control variates once a sensible value of
$\theta^{\star}_n$ is found. Finally, Table \ref{tab:error3datasets}
shows some statistics of the absolute proportional error in the perturbed posterior
in \eqref{eq: perturb post} over $100$ MCMC iterations. It is evident that the perturbed posterior
is very accurate, a result that we also confirm graphically by inspecting
KDE estimates of marginal posteriors (not shown here).
\begin{table}[htbp]
\centering \caption{\emph{Subsampling MCMC vs MCMC}. The table shows the mean, max and
$50,75,99$\% quantiles of the absolute error in \eqref{eq: perturb post}
computed using 100 draws from the perturbed posterior distribution.
The results are shown for the Bankruptcy, HIGGS and Covtype datasets. }

\begin{tabular}{lcccccccccc}
\toprule
 & {\footnotesize{}Mean} &  & {\footnotesize{}Max} &  & {\footnotesize{}$50$\%} &  & {\footnotesize{}$75$\%} &  & {\footnotesize{}$95$\%} & \tabularnewline
\cmidrule{2-2} \cmidrule{4-4} \cmidrule{6-6} \cmidrule{8-8} \cmidrule{10-10}
{\footnotesize{}Bankruptcy} & {\footnotesize{}$1.418\times10^{-6}$} &  & {\footnotesize{}$1.243\times10^{-5}$} &  & {\footnotesize{}$1.246\times10^{-6}$} &  & {\footnotesize{}$1.255\times10^{-6}$} &  & {\footnotesize{}$2.284\times10^{-6}$} & \tabularnewline
{\footnotesize{}HIGGS} & {\footnotesize{}$8.594\times10^{-8}$} &  & {\footnotesize{}$7.104\times10^{-7}$} &  & {\footnotesize{}$7.730\times10^{-8}$} &  & {\footnotesize{}$7.823\times10^{-8}$} &  & {\footnotesize{}$9.072\times10^{-8}$} & \tabularnewline
{\footnotesize{}Covtype} & {\footnotesize{}$5.136\times10^{-8}$} &  & {\footnotesize{}$2.358\times10^{-6}$} &  & {\footnotesize{}$8.207\times10^{-9}$} &  & {\footnotesize{}$8.324\times10^{-9}$} &  & {\footnotesize{}$1.853\times10^{-7}$} & \tabularnewline
\bottomrule
\end{tabular}\label{tab:error3datasets}
\end{table}

\section{Conclusions and Future Research\label{sec:Conclusions}}

We propose a framework for speeding up MCMC by data subsampling for data sets with many independent units.
At each MCMC iteration we use two types of control variates to estimate the log of the likelihood unbiasedly and
efficiently using only a small random fraction of the data. This results in a pseudo marginal sampling scheme with a
slightly perturbed posterior. We also use two correlated sampling schemes to improve the mixing of the Markov chain.
We show that by taking $m = O(n^{\frac12})$, the total variation norm of the error in the perturbed posterior is $O(n^{-2})$ if we have access to the MLE based on all data for constructing the control variates, or $O(n^{-\frac12})$ if the MLE is based on a subset with $\tilde n = O(n^\frac12)$ observations. We also show (more heuristically)  as well as empirically that in regions of high concentration of the  posterior the proportional perturbation error of the posterior is extremely small and much smaller than the corresponding error in the likelihood.
Finally, we document large speed ups relative to MCMC using all the data and show that our method
outperforms other recent subsampling approaches in the literature.

If we change the pseudo marginal sampling scheme to a Metropolis-within-Gibbs one
 where we generate the $u$ conditional on $\theta$ and then $\theta$ conditional on $u$, then
 we can obtain exact derivatives of the log of the estimated likelihood. That means that the subsampling
 approach can use efficient proposals such as those based on Gibbs sampling, Laplace approximations and Langevin diffusions
 and so can readily scale up in terms of the number of unknown parameters.

One immediate application of our methods will be to problems where computing the density of each data unit
is very expensive, although the number of data units is not necessarily large. This may be the case when latent variables are present so the density of each observation is an integral.

\section{Acknowledgments}
We would like to thank the Reviewers and the Associate Editor for helping to improve both the content and the presentation of the paper.
Matias Quiroz and Robert Kohn were partially supported by Australian
Research Council Center of Excellence grant CE140100049. Quiroz was
also partially supported by VINNOVA grant 2010-02635. Mattias Villani
was partially financially supported by Swedish Foundation for Strategic
Research (Smart Systems: RIT 15-0097). Minh-Ngoc Tran was partially
supported by a Business School Pilot Research grant. We thank the
authors in \citet{bardenet2015markov} for making their code publicly
available, which facilitated the comparison against other subsampling
approaches.

\bibliographystyle{apalike}
\addcontentsline{toc}{section}{\refname}\bibliography{ref}

\clearpage 

\renewcommand{\thealgorithm}{S\arabic{algorithm}}
\renewcommand{\theremark}{S\arabic{remark}}
\renewcommand{\theequation}{S\arabic{equation}}
\renewcommand{\thetheorem}{S\arabic{theorem}}
\renewcommand{\thesection}{S\arabic{section}}
\renewcommand{\thepage}{S\arabic{page}}
\renewcommand{\thetable}{S\arabic{table}}
\renewcommand{\thefigure}{S\arabic{figure}}
\setcounter{page}{1}
\setcounter{section}{0}
\setcounter{equation}{0}
\setcounter{algorithm}{0}
\setcounter{table}{0}

Online supplement to \lq Speeding up MCMC by Efficient Data Subsampling\rq

\section{Construction of data expanded control variates\label{AppendixTop:proxies}}

For brevity, this section omits showing the dependence on $n$ for $q$ and
$\ell$.

\subsection{Local data clusters\label{subsec:Local-data-clusters}}

Let $z^{c}$ and $N_{c}$ denote the centroid and the number of observations
in cluster $c$, respectively. Note that $\sum_{c=1}^{K}N_{c}=n$
and we take $K\ll n$. Algorithm \ref{Alg:Cluster} is an easily implemented
clustering algorithm. The maximum distance $\epsilon$ between an
observation and the cluster is a user defined input. The clustering
is a one-time cost whose output can be stored for future use, and
is easily sequentially updated as new data arrives. For models with
a categorical response, we cluster separately for each category (i.e.
$z_{i}=x_{i}$). The radius $\epsilon$ can be chosen by simple trial
and error to roughly target a preferred $K/n$ ratio. Like any clustering
method, Algorithm \ref{Alg:Cluster} eventually suffers from the curse
of dimensionality in large dimensional data spaces: however, high-dimensional
data tends to cluster on a subspace so the effective dimension may
be substantially smaller. Moreover, as discussed in Section~\ref{subsec:Strategy-for-subsampling},
with a reasonably large $p$ we can still allow for sparse clusters
at the cost of having a large variance of our estimator because we
can effectively reduce the $O(m^{-2})$ error by increasing the subsample
size $m$.

\begin{algorithm}[!b]
\caption{Clustering data points within an $\epsilon$-radius ball}
\label{Alg:Cluster}

\tiny{
\begin{algorithmic}[1]
\Procedure{ClusterData}{$y,x,\epsilon$}
\State $z_i \gets (y_i, x_i)^T$
\State $z_{~} \gets (z_1^T, \dots, z_n^T)^T$. \Comment Standardized data.
\vspace{1mm}
\State $I \gets (0, \dots, 0)^T$ \Comment 0 - observation is not clustered.
\vspace{1mm}
\State $(j,k) \gets (0,0)$ \Comment Initialize counters.
\vspace{1mm}
\While{$\sum I_j \neq n$}
\vspace{1mm}
\If{$I_j = 0$} \Comment If not clustered yet.
\vspace{1mm}
\State $C_k \gets \{i; ||z_j-z_i|| \leq \epsilon \}$ \Comment Form cluster within an $\epsilon$-ball.
\vspace{1mm}
\State $N_k \gets |C_k|$
\vspace{1mm}
\State $z^{c_k} \gets \frac{1}{N_k}\sum_{i\in C_k} z_i$ \Comment Create centroid with $N_k$ observations.
\vspace{1mm}
\State $I_{C_k} \gets 1$ \Comment Mark clustered observations.
\vspace{1mm}
\State $k \gets k+1$
\vspace{1mm}
\EndIf
\vspace{1mm}
\State $j \gets j + 1$
\vspace{1mm}
\EndWhile{}
\vspace{1mm}
\State $K \gets k$
\vspace{1mm}
\State \Return $\{z^{c_k}\}_{k=1}^{K}$, $\{C_k\}_{k=1}^{K}$
\EndProcedure
\end{algorithmic}
}
\end{algorithm}

\subsection{Data based control variates}\label{SS: data based control variates}

For notational clarity we consider a univariate response $y$ and
write
\[
\ell(z_{i};\theta)\coloneqq\log p(y_{i}|x_{i},\theta)=\ell_{i}(\theta)
\]
to emphasize that we now consider $\ell_{i}$ as a function of the
data $z_{i}=(y_{i},x_{i})^{T}\in\text{\ensuremath{\mathbb{R}}}^{(p+1)\times1}$
for a given parameter $\theta\in\mathbb{R}^{p}$. Let $C$ denote
the index set of observations within cluster $c$. For any $i\in C$,
a second order Taylor approximation of $\ell(z_{i};\theta)$ around
the centroid $z^{c}$ is
\begin{eqnarray*}
q(z_{i};\theta) & = & \ell(z^{c};\theta)+\triangledown_{z}\ell(z^{c};\theta)^{T}(z_{i}-z^{c})+\frac{1}{2}(z_{i}-z^{c})^{T}H(z^{c};\theta)(z_{i}-z^{c}),
\end{eqnarray*}
where $H(z^{c};\theta)=\triangledown_{z}^{2}\ell(z^{c};\theta)$ is
the Hessian evaluated at $z^{c}$. Note that once $\ell(z^{c};\theta)$
is computed, it is relatively cheap to evaluate $\triangledown_{z}\ell(z^{c};\theta)$
and $H(z^{c};\theta)$ by using the chain rule for differentiation.
The next subsection provides formulas for computing $q=\sum_{i=1}^{n}q(z_{i};\theta)$
at the centroids $\{z^{c_{k}}\}_{k=1}^{K}$, where typically $K\ll n$.

The approximation error is given by the remainder term of the Taylor
series, which depends on the clustering algorithm through
$\epsilon$ in Algorithm~\ref{Alg:Cluster}, and is the maximum
distance between an observation in a cluster and its centroid. The
choice of $\epsilon$ determines how local the approximation is: the
smaller the $\epsilon$ the larger the number of clusters $K$. In
our applications we choose $K$ to optimize the PM sampling efficiency.
If this results in a poor approximation it is compensated by $m$
which reduces the error as $O(m^{-2})$ for fixed $n$.

\subsection{Compact matrix computations}

We now outline how to compute $\sum_{i=1}^{n}q_{i}(\theta)$ by only
scaling quantities evaluated at the centroids.

Let $z^{c_{k}}$ denote the centroid in cluster $c_{k},$ $k=1,\dots,K$.
Let $C_{k}$ denote the index set of observations within cluster $c_{k}$
with $N_{k}=|C_{k}|$. The second order Taylor approximation $\ell(z_{i};\theta)$
in cluster $c_{k}$, for $i\in C_{k}$, is
\begin{eqnarray*}
q(z_{i};\theta) & = & \ell(z^{c_{k}};\theta)+\triangledown_{z}\ell(z^{c_{k}};\theta)^{T}(z_{i}-z^{c_{k}})+\frac{1}{2}(z_{i}-z^{c_{k}})^{T}H(z^{c_{k}};\theta)(z_{i}-z^{c_{k}}).
\end{eqnarray*}
We now derive a compact expression for $\sum_{i=1}^{n}q(z_{i};\theta)$,
i.e.
\begin{equation}
\sum_{k=1}^{K}\sum_{i\in C_{k}}\ell(z^{c_{k}};\theta)+\sum_{k=1}^{K}\sum_{i\in C_{k}}\triangledown_{z}\ell(z^{c_{k}};\theta)^{T}(z_{i}-z^{c_{k}})+\frac{1}{2}\sum_{k=1}^{K}\sum_{i\in C_{k}}(z_{i}-z^{c_{k}})^{T}H(z^{c_{k}};\theta)(z_{i}-z^{c_{k}}).\label{eq:sum_of_q}
\end{equation}
 Note that, within a centroid $c_{k}$, $\ell(z^{c_{k}};\theta),\triangledown_{z}\ell(z^{c_{k}};\theta)$
and $H(z^{c_{k}};\theta)$ are constant. Therefore the first term
in \eqref{eq:sum_of_q} is
\[
\sum_{k=1}^{K}\sum_{i\in C_{k}}\ell(z^{c_{k}};\theta)=\sum_{k=1}^{K}\ell(z^{c_{k}};\theta)\sum_{i\in C_{k}}1=\sum_{k=1}^{K}N_{k}\ell(z^{c_{k}};\theta).
\]
For the middle term in \eqref{eq:sum_of_q}, we have
\[
\sum_{k=1}^{K}\sum_{i\in C_{k}}\triangledown_{z}\ell(z^{c_{k}};\theta)^{T}(z_{i}-z^{c_{k}})=\sum_{k=1}^{K}\triangledown_{z}\ell(z^{c_{k}};\theta)^{T}\sum_{i\in C_{k}}(z_{i}-z^{c_{k}}),
\]
where $\sum_{i\in C_{k}}(z_{i}-z^{c_{k}})\in\mathbb{R}^{(p+1)\times1}$
is the vector sum of the indices in $C_{k}$ for the $k$th centroid,
independent of $\theta$ and hence only needs to be computed once
before the MCMC.

For the last term in \eqref{eq:sum_of_q}, by the definition of the
quadratic form
\begin{eqnarray*}
b_{i}^{T}H^{(k)}b_{i} & = & \sum_{s,t}H_{st}^{(k)}b_{is}b_{it},
\end{eqnarray*}
with $b_{i}=(z_{i}-z^{c_{k}})^{T}\in\mathbb{R}^{(p+1)\times1}$ and
$H^{(k)}=H(z^{c_{k}};\theta)$ we obtain
\begin{eqnarray*}
\sum_{k=1}^{K}\sum_{i\in C_{k}}b_{i}^{T}H^{(k)}b_{i} & = & \sum_{k=1}^{K}\sum_{i\in C_{k}}\sum_{s,t}H_{st}^{(k)}b_{is}b_{it}\\
 & = & \sum_{s,t}\left(\sum_{k=1}^{K}H_{st}^{(k)}\sum_{i\in C_{k}}b_{is}b_{it}\right).
\end{eqnarray*}
Let $B^{(k)}$ be a $\mathbb{R}^{(p+1)\times(p+1)}$ matrix with elements
$\{\sum_{i\in C_{k}}b_{ij}b_{ik}\}_{jk}$. Then
\begin{eqnarray*}
\sum_{k=1}^{K}\sum_{i\in C_{k}}b_{i}^{T}H^{(k)}b_{i} & = & \sum\mathrm{vec}\left(\sum_{k=1}^{K}H^{(k)}\circ B^{(k)}\right),
\end{eqnarray*}
where $\circ$ denotes the Hadamard product (element-wise multiplication)
and the sum without indices is over all elements after vectorization.
$B^{(k)}$ does not depend on $\theta$ so we can compute it before
the MCMC.

We assume that the dominating cost of the MCMC is the density evaluations.
In data sets with a reasonable number of covariates, the term $\sum_{k=1}^{K}\sum_{i\in C_{k}}b_{i}^{T}H^{(k)}b_{i}$
might be costly as it involves $K\text{\ensuremath{\times}(}p+1)^{2}$
summations, which reduces to $K\times\frac{(p+1)(p+2)}{2}$ because
$H$ and \textbf{$B$ }are symmetric. In models where the density
is log-concave (or convex) we have found that evaluating the second
order term in the Taylor approximation for a fixed $\theta$, e.g.
the posterior mode, provides a good approximation.

\subsection{Computing the data expanded control variates for the GLM class}

We now derive the control variates based on expansion around $z$
for the class of Generalized Linear Models (GLM, \citealt{nelder1972glm}).
We emphasize that our method applies much more widely: the only requirement
is that $\ell(z;\theta)$ is twice differentiable with respect to
$z$. We note that categorical variables, either response or covariates,
are considered as continuous in the differentiation.

Consider a univariate GLM
\begin{align*}
p(y|x,\theta) & \coloneqq  h(y)g(\Psi)\exp\left(b(\Psi)T(y)\right),
\end{align*}
with $\E[y|x]\coloneqq\Psi$, with $k(\Psi)=x^{T}\theta$
for an invertible link function $k$.
The log-density as a function of data $z=(y,x)^{T}\in\mathbb{R}^{(p+1)\times1}$
is
\begin{eqnarray*}
\ell(z;\theta) & = & \log(h(y))+\log(g(\Psi))+b(\Psi)T(y)\\
\Psi & = & k^{-1}(x^{T}\theta).
\end{eqnarray*}

To save space, define
\begin{align*}
k^{-1^{\prime}} & \coloneqq  \left.\frac{d}{da}k^{-1}(a)\right|_{a=x^{T}\theta} \quad \text{and} \quad
k^{-1^{\prime\prime}} \coloneqq  \left.\frac{d^{2}}{da^{2}}k^{-1}(a)\right|_{a=x^{T}\theta}.
\end{align*}
The gradient $\triangledown_{z}\ell(z;\theta)$ is the $\mathbb{R}^{(p+1)\times1}$
vector
\begin{eqnarray*}
\left[\begin{array}{c}
\frac{\partial\ell}{\partial y}\\
\frac{\partial\ell}{\partial x}
\end{array}\right] & = & \left[\begin{array}{c}
\frac{h^{\prime}(y)}{h(y)}+b(\Psi)T^{\prime}(y)\\
\left(\frac{g^{\prime}(\Psi)}{g(\Psi)}k^{-1^{\prime}}+b^{\prime}(\Psi)T(y)\right)\theta
\end{array}\right]
\end{eqnarray*}
evaluated at $\Psi=k^{-1}(x^{T}\theta)$, $\theta\in\mathbb{R}^{p\times1}$.
The Hessian $\triangledown_{z}^{2}\ell(z;\theta)$ is the $\mathbb{R}^{(p+1)\times(p+1)}$
matrix with elements
\[
\begin{bmatrix}\frac{\partial^{2}\ell}{\partial y^{2}} & \frac{\partial^{2}\ell}{\partial y\partial x^{T}}\\
\frac{\partial^{2}\ell}{\partial y\partial x} & \frac{\partial^{2}\ell}{\partial x\partial x^{T}}
\end{bmatrix}
\]
where
\begin{eqnarray*}
\frac{\partial^{2}\ell}{\partial y^{2}} & = & \frac{1}{h(y)}\left(h^{\prime\prime}(y)-\frac{h^{\prime}(y)}{h(y)}\right)+b(\Psi)T^{\prime\prime}(y)\\
\frac{\partial^{2}\ell}{\partial y\partial x} & = & \left(b^{\prime}(\Psi)k^{-1^{\prime}}T^{\prime}(y)\right)\theta\\
\frac{\partial^{2}\ell}{\partial x\partial x^{T}} & = & \left(\left(k^{-1^{\prime}}\right)^{2}\frac{1}{g(\Psi)}\left(g^{\prime\prime}(\Psi)-\frac{g^{\prime}(\Psi)}{g(\Psi)}\right)+\frac{g^{\prime}(\text{\ensuremath{\Psi})}}{g(\Psi)}k^{-1^{\prime\prime}}+b^{\prime\prime}(\Psi)k^{-1^{\prime}}T(y)\right)\theta\theta^{T}.
\end{eqnarray*}
We note that even in models with vector valued $\Psi$ (which are
outside the GLM class) it is typically straightforward to derive the
approximation.

\section{Proofs\label{AppendixTop:Proofs}}

\begin{proof}[Proof of Lemma \ref{lem:Unbiased_and_variances}]
The proofs of parts (i) and (ii) are straightforward and omitted. We prove part~(iii).
For $m$-asymptotics, since $u_{i}$'s are iid and $\sigma_{d,n}^{2}<\infty$,
the standard Central Limit Theorem (CLT) gives $\sqrt{m}(\widehat{\mu}_{d,n}-\mu_{d,n})/\sigma_{d,n}\sim\mathcal{N}(0,1)$.
The result for $\widehat{\ell}_{(m,n)}$ follows easily as $n$ is
fixed. For $n$-asymptotics, let $m=Bn^{\alpha}$ for constants $B>0$
and $\alpha>0$ and define $\mathbb{P}_{n}(x)=\Pr\left(\sqrt{B}n^{\alpha/2}\frac{(\widehat{\ell}_{(m,n)}-\ell_{(n)})}{n\sigma_{d,n}}\leq x\right)$.
By the Berry-Esseen theorem \citep{Berry:1941,Esseen:1942}
\begin{eqnarray*}
\left|\mathbb{P}_{n}(x)-\Phi(x)\right| & \leq & \frac{C}{\sqrt{B}n^{\alpha/2}}\frac{\E\left[|d_{u,n}-\mu_{d,n}|^{3}\right]}{\sigma_{d,n}^{3}},\quad\text{where }C\text{ is a constant.}
\end{eqnarray*}
It is straightforward to show that $\frac{\E\left[|d_{u,n}-\mu_{d,n}|^{3}\right]}{\sigma_{d,n}^{3}}=O(1)$
implying a CLT for $\widehat{\ell}_{(m,n)}$ whenever $\gamma>0$.
Proof of part (iv). It is straightforward to show that ${\rm Var} (\wh \sigma^2 _{d,n}) = O(a_n^4) /m $ so that
 ${\rm Var} (\frac{n^2}{2m} \wh \sigma^2 _{d,n}) = \frac{n^4}{m^3} O(a_n^4)=m^{-1}\sigma^4_{LL,m,n}  $ so the result holds as long as $\sigma^2_{LL,m,n}/m = O(1)$.
\end{proof}

Before proving Theorem \ref{thm:maintheorem}, we note that parts (A1)-(A4) of Assumption \ref{Ass2} imply the conditions P1, P2, C1 and C2 in \cite{Chen:1985}, therefore we have the following lemma, which is Lemma 2.1 in \cite{Chen:1985}.

\begin{lemma}\label{Lemma2.1Chen}
Assume that the sequence of the posteriors $\{\pi_{(n)},n=1,2,...\}$ satisfies Part (A1)-(A4) of Assumption \ref{Ass2}, then
\[\lim_{n\to\infty}\pi_{(n)}(\theta^\star_n)|\Sigma_n|^{1/2}\leq (2\pi)^{-p/2}.\]
\end{lemma}

\subsection*{Proof of Theorem \ref{thm:maintheorem}}
We first show that, for each $\theta\in\Theta$,
\begin{equation}
\E\left[\exp\left(\widehat{\ell}_{(m,n)}-\frac{n^{2}}{2m}\widehat{\sigma}_{d,n}^{2}\right)\right]\leq\exp\left(\ell_{(n)}\right)\left(1+O\left(\left(\frac{na_{n}}{m}\right)^{2}\right)\right).\label{eq:main_result}
\end{equation}
The proof first decomposes the LHS of \eqref{eq:main_result}, by
defining \textbf{$\widetilde{\sigma}_{d,n}^{2}\coloneqq\frac{1}{m}\sum_{i=1}^{m}\left(d_{u_{i},n}-\text{\ensuremath{\mu}}_{d,n}\right)^{2}$},
as
\begin{align}
\widehat{\ell}_{(m,n)}-\frac{n^{2}}{2m}\widehat{\sigma}_{d,n}^{2} & =\left(\widehat{\ell}_{(m,n)}-\frac{n^{2}}{2m}\sigma_{d,n}^{2}\right)+\frac{n^{2}}{2m}\left(\sigma_{d,n}^{2}-\widetilde{\sigma}_{d,n}^{2}\right)+\frac{n^{2}}{2m}\left(\widetilde{\sigma}_{d,n}^{2}-\widehat{\sigma}_{d,n}^{2}\right),\label{eq: main_decomposition-1}
\end{align}
and then utilizes the following lemma.

\begin{lemma}\label{lem:Exponential_orders-1}Suppose that $X_{m,n},Y_{m,n}$
and $Z_{m,n}$ are random variables such that
\begin{align*}
\E\left[\exp\left(\lambda X_{m,n}\right)\right] & =1+O(a_{m,n}),\quad\E\left[\exp\left(\lambda Y_{m,n}\right)\right]=1+O(b_{m,n}),\\
\E\left[\exp\left(\lambda Z_{m,n}\right)\right] & =1+O(c_{m,n}),
\end{align*}
where $a_{m,n},b_{m,n}$ and $c_{m,n}$ are $o(1)$ for any fixed
$\lambda$. Then,
\[
\E\left[\exp(X_{m,n}+Y_{m,n}+Z_{m,n})\right]=1+O(a_{m,n})+O(b_{m,n})+O(c_{m,n}).
\]
\end{lemma}
\begin{proof}
Applying H\"{o}lder's inequality twice yields the result.
\end{proof}
We first prove \eqref{eq:main_result} assuming
that $d_{u_{i},n}\sim\mathcal{N}(\mu_{d,n},\sigma_{d,n}^{2})$ to
outline the intuition of the result. Then one technical lemma is
given with the aim to prove the theorem for any $d_{u_{i},n}$ and,
in particular, our theory does not rely on normality of the difference
estimator.
\begin{proof}[Proof of \eqref{eq:main_result} under normality of the $d_{u_{i,n}}$]Since
$\widehat{\ell}_{(m,n)}\sim\mathcal{N}(\ell_{(n)},\frac{n^{2}}{m}\sigma_{d,n}^{2})$,
it follows that
\[
\E\left[\exp\left(\widehat{\ell}_{(m,n)}-\frac{n^{2}}{2m}\sigma_{d,n}^{2}\right)\right]=\exp(\ell_{(n)}),
\]
for the first term in \eqref{eq: main_decomposition-1}. Next,
\begin{eqnarray*}
\E\left[\exp\left(\frac{n^{2}}{2m}\left(\sigma_{d,n}^{2}-\widetilde{\sigma}_{d,n}^{2}\right)\right)\right] & = & \E\left[\exp\left(-\frac{n^{2}}{2m^{2}}\left(\sum_{i=1}^{m}\left(d_{u_{i}}-\mu_{d,n}\right)^{2}-\sigma_{d,n}^{2}\right)\right)\right]\\
 & = & \E\left[\exp\left(-mt\left(\sum_{i=1}^{m}\nu_{i}-1\right)\right)\right],
\end{eqnarray*}
with $t=n^{2}\sigma_{d,n}^{2}/(2m^{2})$ and $\nu_{i}\sim\chi^{2}(1)$.
Since the $\nu_{i}$' are iid and using the mgf we get
\begin{eqnarray*}
\E\left[\exp\left(\frac{n^{2}}{2m}\left(\sigma_{d,n}^{2}-\widetilde{\sigma}_{d,n}^{2}\right)\right)\right] & = & \exp\left(mt\right)\left(\E\left[\exp\left(-t\nu\right)\right]\right)^{m}\\
 & = & \exp\left(mt\right)\left(\left(1+2t\right){}^{-1/2})\right)^{m}\\
 & = & \exp\left(mt\right)\exp\left(-mt+mt^{2}\right)\\
 & = & 1+O\left(\frac{n^{4}a_{n}^{4}}{m^{3}}\right).
\end{eqnarray*}
Finally, consider
\begin{eqnarray*}
\E\left[\exp\left(\frac{n^{2}}{2m}\left(\widetilde{\sigma}_{d,n}^{2}-\widehat{\sigma}_{d,n}^{2}\right)\right)\right] & = & \E\left[\exp\left(\frac{n^{2}}{2m}\left(\widehat{\mu}_{d,n}-\mu_{d,n}\right)^{2}\right)\right]\\
 & = & \E\left[\exp\left(\frac{n^{2}\sigma_{d,n}^{2}}{2m^{2}}\nu\right)\right]\\
 & = & \left(1-\frac{n^{2}\sigma_{d,n}^{2}}{m^{2}}\right)^{-1/2}\\
 & = & 1+O\left(\frac{n^{2}\sigma_{d,n}^{2}}{m^{2}}\right) \\
 & = & 1+O\left(\frac{n^{2}a_{n}^{2}}{m^{2}}\right).
\end{eqnarray*}
The result now follows from Lemma \ref{lem:Exponential_orders-1}.

\end{proof}

For the general proof of \eqref{eq:main_result} without the normality assumption,
we use the following lemma which is an application of Bernstein's
inequality.\begin{lemma} \label{lemm: Bernstein's in equality} Suppose
that $X$ is a random variable such that $|\E[X^{r}]|\leq Bb^{r}$
for some $B>0$ and $b>0$. Then, for $0\leq\lambda<1/b$,
\begin{align}
\log\E[\exp(\lambda X)]\leq & \lambda\E[X]+\frac{1}{2}\lambda^{2}\E[X^{2}]+B(\lambda b)^{3}/(1-\lambda b)\label{eq: ineq1}
\end{align}
and
\begin{align}
\E[\exp(\lambda X)] & \leq1+\lambda\E[X]+\frac{1}{2}\lambda^{2}\E[X^{2}]+B(\lambda b)^{3}/(1-\lambda b)\label{eq: ineq2}
\end{align}
\begin{proof}
\begin{align*}
\log\E[\exp(\lambda X)] & \leq\E[\exp(\lambda X)]-1\leq\E[\lambda X+(\lambda X)^{2}/2+\cdots]\\
 & \leq\lambda\E[X]+\frac{1}{2}\lambda^{2}\E[X^{2}]+B(\lambda b)^{3}/3!+B(\lambda b)^{4}/4!\cdots
\end{align*}
and we obtain inequality \eqref{eq: ineq1}. Inequality~\eqref{eq: ineq2}
follows. \end{proof} \end{lemma}

\begin{proof}[Proof of \eqref{eq:main_result} without normality assumption]

For the first term in \eqref{eq: main_decomposition-1}, define iid
$\xi_{i,m,n}=\frac{n}{m}\left(d_{u_{i}}-\mu_{d,n}\right)$ with $\E[\xi_{i,m,n}]=0$
and $\mathrm{\E}[\xi_{i,m,n}^{2}]=\frac{n^{2}}{m^{2}}\sigma_{d,n}^{2}$,
and write
\[
\widehat{\ell}_{(m,n)}-\frac{n^{2}}{2m}\sigma_{d,n}^{2}=\left(\sum_{i=1}^{m}\xi_{i,m,n}\right)+\ell_{(n)}-\frac{n^{2}}{2m}\sigma_{d,n}^{2}.
\]
Then
\begin{eqnarray*}
\E\left[\exp\left(\widehat{\ell}_{(m,n)}-\frac{n^{2}}{2m}\sigma_{d,n}^{2}\right)\right] & = & \exp(\ell_{(n)}-\frac{n^{2}}{2m}\sigma_{d,n}^{2})\E\left[\exp\left(\sum_{i=1}^{m}\xi_{i,m,n}\right)\right]\\
 & = & \exp(\ell_{(n)}-\frac{n^{2}}{2m}\sigma_{d,n}^{2})\E\left[\exp\left(\xi_{m,n}\right)\right]^{m}.
\end{eqnarray*}
Moreover, since $\left|\E[\xi_{m,n}^{r}]\right|\leq\left|(\frac{n}{m})^{r}\E\left[\left(d_{u}-\mu_{d,n}\right)^{r}\right]\right|\leq\left(\frac{2a_{n}n}{m}\right)^{r}$,
applying Lemma \ref{lemm: Bernstein's in equality}
\begin{align*}
\log\E\left[\exp(\lambda\xi_{m,n})\right] & \leq\frac{1}{2}\left(\lambda\frac{n}{m}\right)^{2}\sigma_{d,n}^{2}+\left(\frac{2\lambda a_{n}n}{m}\right)^{3}/\left(1-\frac{2\lambda a_{n}n}{m}\right)
\end{align*}
for $\lambda<m/(2a_{n}n)$, and we can take $\lambda=1$ for $n$
large enough ($n$-asymptotics) and $m$ large enough for fixed $n$.
Thus,
\[
\E\left[\exp(\xi_{m,n})\right]=\exp\left(\frac{n^{2}\sigma_{d,n}^{2}}{2m^{2}}+O\left(\left(\frac{na_{n}}{m}\right)^{3}\right)\right)
\]
and
\begin{align*}
\E\left[\exp\left(\xi_{m,n}\right)\right]^{m} & =\exp\left(\frac{n^{2}\sigma_{d,n}^{2}}{2m}\right)\exp\left(O\left(\frac{n^{3}a_{n}^{3}}{m^{2}}\right)\right).
\end{align*}
It follows that
\begin{eqnarray}
\E\left[\exp\left(\widehat{\ell}_{(m,n)}-\frac{n^{2}}{2m}\sigma_{d,n}^{2}\right)\right] & = & \exp(\ell_{(n)})\exp\left(O\left(\frac{n^{3}a_{n}^{3}}{m^{2}}\right)\right)\nonumber \\
 & = & \exp(\ell_{(n)})\left(1+O\left(\frac{n^{3}a_{n}^{3}}{m^{2}}\right)\right).\label{eq:Order_term_1}
\end{eqnarray}

For the middle term in \eqref{eq: main_decomposition-1}, define iid
$\xi_{i,m,n}=-\frac{n^{2}}{2m^{2}}\left(\left(d_{u_{i}}-\mu_{d,n}\right)^{2}-\sigma_{d,n}^{2}\right)$
with
\[
\E[\xi_{i,m,n}]=0\quad\text{and }\E[\xi_{i,m,n}^{2}]=\frac{n^{4}}{4m^{4}}\E[(d_{u_{i}}-\mu_{d,n})^{4}-\sigma_{d,n}^{4}]=O\left(\left(\frac{na_{n}}{m}\right)^{4}\right).
\]
 We can show that
\[
\left|\E[\xi_{m,n}^{r}]\right|\leq\left|(\frac{n}{m})^{r}\E\left[\left(d_{u}-\mu_{d,n}\right)^{r}\right]\right|\leq\left(\frac{\sqrt{5}a_{n}n}{m}\right)^{2r},
\]
and applying Lemma \ref{lemm: Bernstein's in equality} we conclude
that
\begin{eqnarray}
\E\left[\exp\left(\frac{n^{2}}{2m}\left(\sigma_{d,n}^{2}-\widetilde{\sigma}_{d,(n)}^{2}\right)\right)\right] & = & 1+O\left(\frac{n^{4}a_{n}^{4}}{m^{3}}\right).\label{eq:order_term_2}
\end{eqnarray}
Finally, consider the last term in \eqref{eq: main_decomposition-1}
and let
\begin{eqnarray*}
\xi_{n,m} & = & \frac{n^{2}}{2m}\left(\widetilde{\sigma}_{d,n}^{2}-\widehat{\sigma}_{d,n}^{2}\right)\\
 & = & \frac{n^{2}}{2m}\left(\widehat{\mu}_{d,n}-\mu_{d,n}\right)^{2}\\
 & = & \frac{n^{2}}{2m}\left(\overline{X}\right)^{2},
\end{eqnarray*}
where $\overline{X}=\frac{1}{m}\sum_{i=1}^{m}X_{i}$, $X_{i}=d_{u_{i},n}-\mu_{d,n}$
and
\[
\E[X_{i}]=0,\quad\E[X_{i}^{2}]=\sigma_{d,n}^{2}.
\]
Note that $\left|X_{i}\right|\leq a_{n}$ so that $\left|\overline{X}\right|\leq a_{n}$
and hence $\E\left[\left|\overline{X}\right|^{r}\right]\leq a_{n}^{r}$
for $r\geq1$. Therefore,
\[
\E\left[\xi_{n,m}^{r}\right]=\left(\frac{n^{2}}{2m}\right)^{r}\E\left[\overline{X}^{2r}\right]=O\left(\left(\frac{n^{2}a_{n}^{2}}{m^{2}}\right)^{r}\right)=O\left(\left(\frac{na_{n}}{m}\right)^{2r}\right),
\]
and it follows by Lemma \ref{lemm: Bernstein's in equality} that
\begin{eqnarray}
\E\left[\exp\left(\frac{n^{2}}{2m}\left(\sigma_{d,n}^{2}-\widetilde{\sigma}_{d,(n)}^{2}\right)\right)\right] & = & 1+O\left(\frac{n^{2}a_{n}^{2}}{m^{2}}\right).\label{eq:order_term_3}
\end{eqnarray}
Applying Lemma \ref{lem:Exponential_orders-1} on \eqref{eq:Order_term_1},
\eqref{eq:order_term_2}, \eqref{eq:order_term_3} and concluding
that the slowest decaying term is $O\left(\frac{n^{2}a_{n}^{2}}{m^{2}}\right)$
proves \eqref{eq:main_result}.
\end{proof}

We now prove the main results of Theorem \ref{thm:maintheorem}. \begin{proof}[Proof of part (i)-(ii) of Theorem 1] By \eqref{eq:main_result},
\[
\left|{L_{(m,n)}(\theta)-L_{(n)}(\theta)}\right|\leq L_{(n)}(\theta) O\left(\left(\frac{na_{n}(\theta)}{m}\right)^{2}\right).
\]
That is, there exists an $M_1>0$ such that
\[
\left|{L_{(m,n)}(\theta)-L_{(n)}(\theta)}\right|\leq M_1\frac{n^2}{m^2}L_{(n)}(\theta)a_{n}^2(\theta).
\]
Hence,
\begin{eqnarray}\label{eq:this_equation}
\left|\overline{L}_{(m,n)}-\overline{L}_{(n)}\right| & \leq & \int\left|L_{(m,n)}(\theta)-L_{(n)}(\theta)\right|p_{\Theta}(\theta)d\theta.\notag\\
 & \leq & M_1\frac{n^2}{m^2}\overline{L}_{(n)} \E_{\pi_{(n)}}\left(a_{n}^2(\theta)\right).
\end{eqnarray}
Let $x = \Sigma^{-1/2}_n(\theta - \theta_n^\star)$ and recall that there exists $M_2>0$,
\[a_n(\theta)\leq M_2\|\theta-\theta_n^\star\|^3=M_2\|\Sigma_n^{1/2}x\|^3\leq M_2\|\Sigma_n\|^{3/2}\|x\|^3.\]
For any $\delta > 0 $, write
\begin{align*}
\int a_n(\theta)^2 \pi_{(n)}(\theta) d \theta & = I_{n,\delta} + II_{n,\delta}
\end{align*}
where
\begin{align*}
I_{n,\delta}& = \int_{|| \theta - \theta_n^\star|| < \delta }a_n(\theta)^2 \pi_{(n)}(\theta) d \theta\quad \text{and}
\quad II_{n,\delta} = \int_{|| \theta - \theta_n^\star|| \geq  \delta }a_n(\theta)^2 \pi_{(n)}(\theta) d \theta
\end{align*}
Consider first $I_{n,\delta}$. We have
\begin{align*}
\log \pi_{(n)} (\theta) - \log \pi_{(n)} (\theta_n^\star) & = \frac{1}{2} (\theta - \theta_n^\star)^{T} \Delta_n(\wt \theta) (\theta - \theta_n^\star),
\end{align*}
where $\tilde \theta$ lies between $\theta$ and $\theta_n^\star$. From part (A4) of Assumption 1 and a little algebra,
for any $\epsilon > 0 $, there is a $\delta_\epsilon> 0 $ and a positive integer $N_{1,\epsilon} $ such that
for $||\theta - \theta_n^\star||< \delta_\epsilon$ and $n \geq N_{1,\epsilon}$
\begin{align*}
\frac12 (\theta - \theta_n^\star)^{T}  \Delta_n(\wt \theta) (\theta - \theta_n^\star)& \leq
- \frac12 (\theta - \theta_n^\star)^{T}  \Sigma_n^{-1}(\wt \theta) (\theta - \theta_n^\star)
+\frac12 (\theta - \theta_n^\star)^{T}  A(\epsilon)\Sigma_n^{-1} (\theta - \theta_n^\star)\\
& = -\frac12 x^{T}x + \frac12 x^{T}\Sigma_n^{\frac12}A(\epsilon)\Sigma_n^{-\frac12}x,
\end{align*}
where $x = \Sigma_n^{-\frac12}(\theta - \theta_n^\star)$. We note that $d\theta  = |\Sigma_n|^\frac12 dx$,
 $\pi_{(n)}(\theta_n^\star) |\Sigma_n|^\frac12 \leq (2\pi)^{-p/2} $ (by Lemma 5)
\begin{align*}
\{\theta : ||\theta - \theta_n^\star||< \delta_\epsilon\}\subset \{x :||x||\leq \delta_\epsilon /{\ul \sigma_n^\frac12}\},
\end{align*}
where $\ul \sigma_n$ is the smallest eigenvalue of $\Sigma_n$. Let $B_n(\epsilon)=\Sigma_n^{\frac12}A(\epsilon)\Sigma_n^{-\frac12}$. Then,
\begin{align*}
I_{n,\delta_\epsilon} & \leq O(1/n^3) \int_{||x|| < \delta_\epsilon/{\ul \sigma_n} } ||x||^6
\exp\left (-\frac12 x^{T}(I - B_n(\epsilon)x \right ) dx \\
& \leq O(1/n^3) \int ||x||^6
\exp\left (-\frac12 x^{T}(I - B_n(\epsilon)x \right ) dx = O(1/n^3).
\end{align*}
Consider now $II_{n,\delta}$. We have for $|| \theta- \theta_n^\star|| > \delta_\epsilon$
\begin{align*}
\pi_{(n)}(\theta)d\theta  & \leq \pi_{(n)}(\theta_n^\star)\exp \left ( -c (x^{T} x)^\kappa \right ) |\Sigma_n| ^\frac12 dx \leq (2\pi)^{-p/2}\exp \left ( -c (x^{T} x)^\kappa \right ) dx  \end{align*}
Hence, for some $M> 0$ and independent of $\theta$ and $n$,
\begin{align*}
II_{n,\delta} & \leq M \int_{||\Sigma_n^\frac12 x|| > \delta_\epsilon} ||\Sigma_n^\frac12 x||^6
\exp \left ( -c (x^{T} x)^\kappa \right )\\
& \leq || \Sigma_n||^3\int ||x||^6 \exp \left ( -c (x^{T} x)^\kappa \right ) = O(1/n^3).
\end{align*}
We have therefore shown that $E_{\pi_{(n)}}(a_n^2(\theta)) = O(1/n^3)$.

Thus, by \eqref{eq:this_equation},
\begin{equation}\label{eq:L_inequality}
\frac{\left|\overline{L}_{(m,n)}-\overline{L}_{(n)}\right|}{\overline{L}_{(n)}} = O\left(\frac{1}{nm^2}\right),
\end{equation}
which implies that
\[\frac{\overline{L}_{(m,n)}}{\overline{L}_{(n)}} = 1+O\left(\frac{1}{nm^2}\right),\;\;\text{and}\;\;\frac{\overline{L}_{(n)}}{\overline{L}_{(m,n)}} = 1+O\left(\frac{1}{nm^2}\right).\]
Now, notice that
\begin{eqnarray*}
\overline{\pi}_{(m,n)}(\theta)-\pi_{(n)}(\theta) & = & L_{(m,n)}(\theta)p_{\Theta}(\theta)/\overline{L}_{(m,n)}-L_{(n)}(\theta)p_{\Theta}(\theta)/\overline{L}_{(n)}\\
 & = & \left(L_{(m,n)}(\theta)-L_{(n)}(\theta)\right)\frac{p_{\Theta}(\theta)}{\overline{L}_{(n)}}\frac{\overline{L}_{(n)}}{\overline{L}_{(m,n)}}\\
 &  & -L_{(n)}(\theta)p_{\Theta}(\theta)\left(\frac{1}{\overline{L}_{(n)}}-\frac{1}{\overline{L}_{(m,n)}}\right).
\end{eqnarray*}
Hence,
\begin{eqnarray*}
\left|\overline{\pi}_{(m,n)}(\theta)-\pi_{(n)}(\theta)\right| & \leq & \left|L_{(m,n)}(\theta)-L_{(n)}(\theta)\right|\frac{p_{\Theta}(\theta)}{\overline{L}_{(m,n)}}+
\frac{\left|\overline{L}_{(m,n)}-\overline{L}_{(n)}\right|}{\overline{L}_{(m,n)}}\pi_{(n)}(\theta).
\end{eqnarray*}
By \eqref{eq:this_equation} and \eqref{eq:L_inequality},
\begin{eqnarray*}
\int_{\Theta}\left|\overline{\pi}_{(m,n)}-\pi_{(n)}(\theta)\right|d\theta & \leq & \frac{1}{\overline{L}_{(m,n)}}\int\left|L_{(m,n)}(\theta)-L_{(n)}(\theta)\right|p_{\Theta}(\theta)d\theta+
\frac{\left|\overline{L}_{(m,n)}-\overline{L}_{(n)}\right|}{\overline{L}_{(m,n)}}\\
&\leq&\frac{\overline{L}_{(n)}}{\overline{L}_{(m,n)}} O\left(\frac{1}{nm^2}\right)+\left|\frac{\overline{L}_{(n)}}{\overline{L}_{(m,n)}}-1\right|\\
&=&O\left(\frac{1}{nm^2}\right),
\end{eqnarray*}
which completes part (i).

To prove part (ii), we have that
\begin{eqnarray*}
\left|{\E_{\overline{\pi}_{(m,n)}}[h(\theta)]-\E_{\pi_{(n)}}[h(\theta)]}\right|&\leq&\int \left|h(\theta)\right|\left|\overline{\pi}_{(m,n)}(\theta)-\pi_{(n)}(\theta)\right|d\theta\\
&\leq&M_1\frac{n^2}{m^2}\frac{\overline{L}_{(n)}}{\overline{L}_{(m,n)}}\int a_n^2(\theta)|h(\theta)|\pi_{(n)}(\theta)d\theta+\\
&&+\frac{\left|\overline{L}_{(m,n)}-\overline{L}_{(n)}\right|}{\overline{L}_{(m,n)}}\int |h(\theta)|\pi_{(n)}(\theta)d\theta\\
&\leq&M_1\frac{n^2}{m^2}\frac{\overline{L}_{(n)}}{\overline{L}_{(m,n)}}\left(\E_{\pi_{(n)}}\left(a_n^4(\theta)\right)\right)^{1/2}\left(\E_{\pi_{(n)}}\left(h^2(\theta)\right)\right)^{1/2}+\\
&&+\frac{\left|\overline{L}_{(m,n)}-\overline{L}_{(n)}\right|}{\overline{L}_{(m,n)}}\left(\E_{\pi_{(n)}}\left(h^2(\theta)\right)\right)^{1/2}.
\end{eqnarray*}
The second term dominates the first term and is of order $O\left(\frac{1}{nm^2}\right)$, which proves the result.\end{proof}

  \begin{proof} [Proof of Corollary~\ref{corr: cv implications}]
  \begin{align*}
 || \theta - \theta_{\wt n}^\star|| & = || \Sigma_n^\frac12 x +\theta_n^\star - \theta_{\wt n}^\star||\\
  & \leq ||\Sigma_n^\frac12|| ||x|| + |O(\wt n^{-\frac12} )|
   \leq \wt n^{-\frac12} |O(1)|\left ( \frac{\wt n^{\frac12}}{n^\frac12} ||x|| + |O(1)| \right )
\end{align*}
Hence,
\begin{align*}
a_n(\theta)&\leq O(1) || \theta - \theta_{\wt n}^\star ||^3
 \leq | O(1)| (\wt n^{- \frac32}) ( ||x|| +|O(1)|)^3
\end{align*}
The rest of the proof is now similar to that of Theorem~\ref{thm:maintheorem}.
  \end{proof}

\section{Checking the assumptions for generalized linear models\label{app:glms}}
Finally, we show that Assumption~\ref{ass: assump for mle} holds for generalized linear models.
Section~5 of Chen~(198) shows how the assumptions of his Theorem 2.1 apply to an exponential family when the prior is conjugate. We use a similar approach for the case when
\begin{align*}
\ell_{(n)}(\theta)& =\log p(y_i|\theta) \propto S_i^{T} \theta - B_i(\theta)
\end{align*}
where the proportionality sign means that there may be an extra term on the right that does not depend on $\theta$.
We assume that $S_i$ does not depend on $\theta$, the third derivative of $B_i$ is continuous, and
${\ddot B}_i(\theta) = \partial^2 B_i(\theta)/\partial \theta \partial \theta^{T}$ is positive semi-definite.
Let
$S_{(n)} = \sum_{i=1}^n S_i/n$ and $B_{(n)}(\theta) = \sum_{i=1}^n B_i(\theta)/n$. We will also assume that $\ddot B_{(n)}(\theta)$ is positive definite for all $\theta$ for $n \geq n_1$ say. Then
$\ell_{(n)}(\theta) = n(S_{(n)}^{\transp} \theta - B_{(n)}(\theta) ) $, the MLE $\theta_n^\star$ is unique and satisfies
$S_{(n)} = \partial B_{(n)}(\theta_n^\star)/\partial \theta $. $\Delta_n(\theta) = - n B_{(n)}(\theta)$ is negative definite for all $\theta$. Then, parts (A1)-(A3) of Assumption~\ref{Ass2} are satisfied. It is also clear that part (A4) is satisfied. We can also use the same approach as in \cite{Chen:1985} to show that part (A5) also holds with $\kappa = 1/2$. If $\ul \sigma_n(\theta) $ is the minimum eigenvalue of $B_{(n)}(\theta)$ and $\sup_{\theta \in \Theta} \ul \sigma_n(\theta) > 0 $ then (A5) also holds with $\kappa = 1$.

The results also generalize in a straightforward way to the case $\ell_i(\theta) \propto  -k_i(x_i^{T} \theta) - B_i(\theta) $, where $k_i(t)$ has continuous third derivative and ${\ddot k_i}(t) \geq 0 $ for all $t$.

\end{document}